\documentclass[twocolumn]{aastex62}
\usepackage{lineno}

\usepackage{newtxtext,newtxmath}
\UseRawInputEncoding

\usepackage{amsmath}	
\usepackage{latexsym}
\usepackage{amssymb}	
\usepackage{gensymb}
\usepackage{flafter}
\usepackage{appendix}
\usepackage{natbib}
\graphicspath{{./}{figures/}}
\newcommand\target{{MAXI~J1820+070}}
\newcommand\hxmt{{\it Insight}-HXMT}

\graphicspath{{./}{figures/}}

\received{January 1, 2018}
\revised{January 7, 2018}
\accepted{\today}
\submitjournal{ApJ}

\shorttitle{The accretion flow geometry of MAXI J1820+070}

\begin{document}

\title{The accretion flow geometry of MAXI J1820+070 through broadband noise research with \hxmt}

\correspondingauthor{Zi-Xu Yang}
\email{yangzx@ihep.ac.cn}
\correspondingauthor{Liang, Zhang}
\email{zhangliang@ihep.ac.cn}
\correspondingauthor{J.L. Qu}
\email{qujl@ihep.ac.cn}

\author{Zi-Xu Yang}
\affil{Key Laboratory for Particle Astrophysics, Institute of High Energy Physics, Chinese Academy of Sciences, 19B Yuquan Road, Beijing 100049, People's Republic of China}
\affil{University of Chinese Academy of Sciences, Chinese Academy of Sciences, Beijing 100049, People's Republic of China}

\author{Liang Zhang}
\affil{Key Laboratory for Particle Astrophysics, Institute of High Energy Physics, Chinese Academy of Sciences, 19B Yuquan Road, Beijing 100049, People's Republic of China}

\author{Qing-Cui Bu}
\affil{Institut f\"ur Astronomie und Astrophysik, Kepler Center for Astro and Particle Physics, Eberhard Karls Universit\"at, Sand 1, 72076 T\"ubingen, Germany」}

\author{Yue Huang}
\affil{Key Laboratory for Particle Astrophysics, Institute of High Energy Physics, Chinese Academy of Sciences, 19B Yuquan Road, Beijing 100049, People's Republic of China}

\author{He-Xin Liu}
\affil{Key Laboratory for Particle Astrophysics, Institute of High Energy Physics, Chinese Academy of Sciences, 19B Yuquan Road, Beijing 100049, People's Republic of China}
\affil{University of Chinese Academy of Sciences, Chinese Academy of Sciences, Beijing 100049, People's Republic of China}

\author{Wei Yu}
\affil{Key Laboratory for Particle Astrophysics, Institute of High Energy Physics, Chinese Academy of Sciences, 19B Yuquan Road, Beijing 100049, People's Republic of China}
\affil{University of Chinese Academy of Sciences, Chinese Academy of Sciences, Beijing 100049, People's Republic of China}

\author{P.J. Wang}
\affil{Key Laboratory for Particle Astrophysics, Institute of High Energy Physics, Chinese Academy of Sciences, 19B Yuquan Road, Beijing 100049, People's Republic of China}
\affil{University of Chinese Academy of Sciences, Chinese Academy of Sciences, Beijing 100049, People's Republic of China}

\author{L. Tao}
\affil{Key Laboratory for Particle Astrophysics, Institute of High Energy Physics, Chinese Academy of Sciences, 19B Yuquan Road, Beijing 100049, People's Republic of China}
\affil{University of Chinese Academy of Sciences, Chinese Academy of Sciences, Beijing 100049, People's Republic of China}

\author{J.L. Qu}
\affil{Key Laboratory for Particle Astrophysics, Institute of High Energy Physics, Chinese Academy of Sciences, 19B Yuquan Road, Beijing 100049, People's Republic of China}
\affil{University of Chinese Academy of Sciences, Chinese Academy of Sciences, Beijing 100049, People's Republic of China}

\author{S. Zhang}
\affil{Key Laboratory for Particle Astrophysics, Institute of High Energy Physics, Chinese Academy of Sciences, 19B Yuquan Road, Beijing 100049, People's Republic of China}
\affil{University of Chinese Academy of Sciences, Chinese Academy of Sciences, Beijing 100049, People's Republic of China}

\author{S.N. Zhang}
\affil{Key Laboratory for Particle Astrophysics, Institute of High Energy Physics, Chinese Academy of Sciences, 19B Yuquan Road, Beijing 100049, People's Republic of China}
\affil{University of Chinese Academy of Sciences, Chinese Academy of Sciences, Beijing 100049, People's Republic of China}

\author{X. Ma}
\affil{Key Laboratory for Particle Astrophysics, Institute of High Energy Physics, Chinese Academy of Sciences, 19B Yuquan Road, Beijing 100049, People's Republic of China}
\affil{University of Chinese Academy of Sciences, Chinese Academy of Sciences, Beijing 100049, People's Republic of China}

\author{L.M. Song}
\affil{Key Laboratory for Particle Astrophysics, Institute of High Energy Physics, Chinese Academy of Sciences, 19B Yuquan Road, Beijing 100049, People's Republic of China}
\affil{University of Chinese Academy of Sciences, Chinese Academy of Sciences, Beijing 100049, People's Republic of China}

\author{S.M. Jia}
\affil{Key Laboratory for Particle Astrophysics, Institute of High Energy Physics, Chinese Academy of Sciences, 19B Yuquan Road, Beijing 100049, People's Republic of China}
\affil{University of Chinese Academy of Sciences, Chinese Academy of Sciences, Beijing 100049, People's Republic of China}

\author{M.Y. Ge}
\affil{Key Laboratory for Particle Astrophysics, Institute of High Energy Physics, Chinese Academy of Sciences, 19B Yuquan Road, Beijing 100049, People's Republic of China}
\affil{University of Chinese Academy of Sciences, Chinese Academy of Sciences, Beijing 100049, People's Republic of China}

\author{Q.Z. Liu}
\affil{Purple Mountain Observatory, Chinese Academy of Sciences, Nanjing 210034, People's Republic of China}

\author{J.Z. Yan}
\affil{Purple Mountain Observatory, Chinese Academy of Sciences, Nanjing 210034, People's Republic of China}

\author{D.K. Zhou}
\affil{Key Laboratory for Particle Astrophysics, Institute of High Energy Physics, Chinese Academy of Sciences, 19B Yuquan Road, Beijing 100049, People's Republic of China}
\affil{University of Chinese Academy of Sciences, Chinese Academy of Sciences, Beijing 100049, People's Republic of China}

\author{T.M. Li}
\affil{Key Laboratory for Particle Astrophysics, Institute of High Energy Physics, Chinese Academy of Sciences, 19B Yuquan Road, Beijing 100049, People's Republic of China}
\affil{University of Chinese Academy of Sciences, Chinese Academy of Sciences, Beijing 100049, People's Republic of China}

\author{B.Y. Wu}
\affil{Key Laboratory for Particle Astrophysics, Institute of High Energy Physics, Chinese Academy of Sciences, 19B Yuquan Road, Beijing 100049, People's Republic of China}
\affil{University of Chinese Academy of Sciences, Chinese Academy of Sciences, Beijing 100049, People's Republic of China}

\author{X.Q. Ren}
\affil{Key Laboratory for Particle Astrophysics, Institute of High Energy Physics, Chinese Academy of Sciences, 19B Yuquan Road, Beijing 100049, People's Republic of China}
\affil{University of Chinese Academy of Sciences, Chinese Academy of Sciences, Beijing 100049, People's Republic of China}

\author{R.C. Ma}
\affil{Key Laboratory for Particle Astrophysics, Institute of High Energy Physics, Chinese Academy of Sciences, 19B Yuquan Road, Beijing 100049, People's Republic of China}
\affil{University of Chinese Academy of Sciences, Chinese Academy of Sciences, Beijing 100049, People's Republic of China}

\author{Y.X. Zhang}
\affil{Kapteyn Astronomical Institute, University of Groningen, Postbus 800, 9700 AV Groningen, The Netherlands}

\author{Y.C. Xu}
\affil{Key Laboratory for Particle Astrophysics, Institute of High Energy Physics, Chinese Academy of Sciences, 19B Yuquan Road, Beijing 100049, People's Republic of China}
\affil{University of Chinese Academy of Sciences, Chinese Academy of Sciences, Beijing 100049, People's Republic of China}

\author{B.Y. Ma}
\affil{Key Laboratory for Particle Astrophysics, Institute of High Energy Physics, Chinese Academy of Sciences, 19B Yuquan Road, Beijing 100049, People's Republic of China}
\affil{University of Chinese Academy of Sciences, Chinese Academy of Sciences, Beijing 100049, People's Republic of China}

\author{Y.F. Du}
\affil{Key Laboratory for Particle Astrophysics, Institute of High Energy Physics, Chinese Academy of Sciences, 19B Yuquan Road, Beijing 100049, People's Republic of China}
\affil{University of Chinese Academy of Sciences, Chinese Academy of Sciences, Beijing 100049, People's Republic of China}

\author{Y.C. Fu}
\affil{Key Laboratory for Particle Astrophysics, Institute of High Energy Physics, Chinese Academy of Sciences, 19B Yuquan Road, Beijing 100049, People's Republic of China}
\affil{University of Chinese Academy of Sciences, Chinese Academy of Sciences, Beijing 100049, People's Republic of China}

\author{Y.X. Xiao}
\affil{Key Laboratory for Particle Astrophysics, Institute of High Energy Physics, Chinese Academy of Sciences, 19B Yuquan Road, Beijing 100049, People's Republic of China}
\affil{University of Chinese Academy of Sciences, Chinese Academy of Sciences, Beijing 100049, People's Republic of China}



\begin{abstract}
Here we present a detailed study of the broadband noise in the power density spectra of the black hole X-ray binary MAXI J1820+070 during the hard state of its 2018 outburst, using the Hard X-ray Modulation Telescope (\emph{Insight}-HXMT) observations. The broadband noise shows two main humps, which might separately correspond to variability from a variable disk and two Comptonization regions. We fitted the two humps with multiple Lorentzian functions and studied the energy-dependent properties of each component up to 100--150 keV and their evolution with spectral changes. The lowest frequency component is considered as the sub-harmonic of QPO component and shows different energy dependence compared with other broadband noise components. We found that although the fractional rms of all the broadband noise components mainly decrease with energy, their rms spectra are different in shape. Above $\sim$ 20--30 keV, the characteristic frequencies of these components increase sharply with energy, meaning that the high-energy component is more variable on short timescales. Our results suggest that the hot inner flow in MAXI J1820+070 is likely to be inhomogeneous. We propose a geometry with a truncated accretion disk, two Comptonization regions.

\end{abstract}



\section{Introduction} \label{sec:intro}

Black hole X-ray binaries (BHXBs) in a complete outburst usually show a counter-clockwise `q-shaped' evolution pattern in the hardness-intensity diagram (HID). The different branches of HID correspond to different spectral and temporal states \citep[see][for reviews]{2006ARA&A..44...49R,2016ASSL..440...61B}.
At the beginning of an outburst, the source is observed in a hard state (HS), where its energy spectrum is dominated by a hard power-law component with a photon index of $\sim$1.5--1.7. 
%
%
After leaving the HS, its spectrum becomes softer. A BHXB at this stage usually goes through an intermediate state (IMS) and then reaches a soft state (SS) near the outburst peak. In the IMS, both a soft component and a hard non-thermal component contribute significantly to the energy spectrum. 
%
%
While in the SS, the spectrum is dominated by a thermal component from an accretion disk \citep{1973A&A....24..337S} with a weak power-law tail. 
During the outburst decay, a source generally follows a reverse order and evolves back to quiescence.

BHXBs typically show fast X-ray variability on a wide range of timescales \citep{2014SSRv..183...43B,2019NewAR..8501524I}. Fourier power density spectrum (PDS) serves as a powerful tool to study the fast X-ray variability. The typical power spectrum of BHXBs in hard state usually includes QPO and broadband noise. According to a different quality factor Q (Q is defined as the ratio of central frequency and width), we divide them into QPO (Q > 2) and broadband noise (Q < 2) components in LHS and IMS \citep{2002ApJ...572..392B}. According to different central frequencies and quality factors etc., QPO can be divided into A, B and C types. In low/hard state and hard immediate state, type-C QPO is believed to come from Lense-Thirring precession based on truncated disk/hot inner flow model \citep{2006ApJ...642..420S,2009MNRAS.397L.101I}. For other broadband continuum components which represent fast aperiodic variability, several models have been presented to explain this noise component, including shot noise model, coronal flare model and fluctuation propagation model \citep{1972ApJ...174L..35T,1981ApJ...246..494N,1990A&A...227L..33B,1994ApJ...435L.125M,1996ApJ...469L.109S}. Considering that shot noise model predicts a stationary power spectrum and cannot produce a linear rms-flux relation for different timescale, it is not accepted to explain the broadband noise \citep{2012Ap&SS.341..383L}. Furthermore, \citet{2004MNRAS.347L..61U} showed that rms-flux relation in the accreting millisecond pulsar SAX J1808.4-3658 is coupled with the 401 Hz pulsation. This relation put strict constrain on the origin of rms-flux relation from magnetic caps of the neutron star, which means that the linear relation does not favor the coronal flare model for the X-ray variability. Up to now, fluctuation propagation model is widely accepted because it naturally explains the rms-flux relation for different timescale which is common in X-ray binaries \citep{2001MNRAS.323L..26U,2002PASJ...54L..69N,2004MNRAS.347L..61U,2004A&A...414.1091G,2005MNRAS.359..345U,2010SCPMA..53S..86L,2012Ap&SS.341..383L}. In the fluctuation propagation model \citep{1997MNRAS.292..679L,2012MNRAS.419.2369I,2016AN....337..385I,2019MNRAS.486.4061M}, the broadband noise components are believed to break down at local viscous frequency $f_{\rm visc} \propto 1/R^2$ in power spectrum with Lorentzian shape $1/ (1+ (f/f_{\rm visc} (R))^2)$ in PDS \citep{2014MNRAS.440.2882R,2016AN....337..524R,2016MNRAS.462.4078R,2016AN....337..385I,2017MNRAS.472.3821R,2017MNRAS.469.2011R,2021MNRAS.504..469T}. Perturbation occurs at each radius of the accretion flow, but the fluctuation from the outer region will modulate the inner region because the inward motion of accretion flow. This is the reason why we call it the fluctuation propagation model. \citet{2014MNRAS.440.2882R} applied for the first time PROPFLUC on the BHB MAXI J1543-564 fitting the single-hump power spectrum in a single energy band.  After that, \citet{2016MNRAS.462.4078R} applied PROPFLUC on BHB MAXI J1659-152 using for the first time the hypothesis of fluctuations stirred up and propagating from the disc to hot flow. They fitted simultaneously the power spectra in two energy bands and cross-spectra between two bands. \citet{2017MNRAS.469.2011R} further updated PROPFLUC by introducing the hypothesis of extra variability in the hot flow, damping and different propagation speeds of the fluctuations. \citet{2017MNRAS.472.3821R} modelled the power spectra, time lags and coherence in hard and soft states of Cyg X-1. \citet{2018MNRAS.474.2259M} considered more realistically both forward and backward propagation for the first time and find propagating fluctuations also produce soft lags at high frequency as the reflection process does by numerical simulations. \citet{2018MNRAS.473.2084M,2018MNRAS.480.4040M} built a spectral-timing model to explain the energy dependence of power spectra and phase lag spectra with two Comptonization zones basing on fluctuation propagating model. 

\citet{2015MNRAS.452.3666S}  showed a noise component with a characteristic
frequency above 1 Hz in the hard energy band (4--8 keV) and the same component
at a lower frequency in the soft band (1--2 keV) in a large BHXB sample. The dependence was interpreted as a hint that the soft photons originating in the outer region of the Comptonizing corona whereas hard band locates inner region. But due to the detector energy band limit, only the energy band below 10 keV was implemented \citep{2015MNRAS.452.3666S}. Apart from the energy dependence of characteristic frequency, in LHS, the noise is also slightly stronger at lower energy. The fractional rms of noise as a function of energy is flat or decreases by a few percent from 2 to 20 keV \citep{2011BASI...39..409B}. In previous studies, some authors \citep{2013ApJ...770..135Y,2014MNRAS.441.1177S} also investigated the energy dependence of power spectra but  is limited to narrow energy range. It is necessary to provide more information about the high energy dependence of broadband noise. As a result, we present a wider range band energy band dependence in more detail with the help of \emph{Insight}-HXMT LE, ME and HE detectors for the first time.

\target\ is a new X-ray transient discovered on 2018 March 11 by MAXI/GSC \citep{2009PASJ...61..999M,2018ATel11399....1K}. 
Optical follow-up observations identified a optical counterpart coinciding with ASASSN-18ey \citep{2018ATeL11400....1D}. \citet{2019ApJ...882L..21T} derived a mass function $f(M)=5.18 \pm 0.15~{\rm M}_{\sun}$, dynamically confirming the black hole nature of the source. \citet{2020ApJ...893L..37T} estimated the mass of the black hole to be $5.73<M({\rm M}_{\sun})<8.34$ under 95\% confidence level limits. A precise distance of $2.96 \pm 0.33$ kpc was obtained from radio parallax \citep{2020MNRAS.493L..81A}, with a jet inclination angle $ i = 63 \pm 3 \degree$. The similar distance result $D=2.81^{+0.70}_{-0.39}$ kpc was obtained by \textit{Gaia} EDR3 parallax measurement \citep{2021AJ....161..147B}. By fitting the temperature and radius of the donor, \citet{2022arXiv220113201M} also constrained the distance at $D \approx 3 $kpc. \target\ is likely to harbor a slowly spinning black hole. \citet{2021MNRAS.504.2168G} constrained the spin of the black hole to be $a_{*} = 0.2_{-0.3}^{+0.2}$ by fitting the \hxmt\ broadband spectra. A similar low spin result $a_{*} = 0.14{\pm 0.09}$ was obtained by \citet{2021ApJ...916..108Z}.

During the outburst, \target\ stayed in the HS for almost 4 months from 2018 March to early 2018 July.
The unchanging shape of the Fe line profile \citep{2019MNRAS.490.1350B}, together with the shortening thermal reverberation lags, suggest that the HS evolution is driven by the changes of the corona, rather than the disk \citep{2019Natur.565..198K}.
However, several recent results are inconsistent with this picture and support a truncated disk geometry \citep{2021A&A...654A..14D,2021MNRAS.506.2020D}.
%
%
Type-C QPOs were detected in Optical and X-ray wavelengths \citep{2018ATel11591....1Y,2018ATel11723....1Z,2018ATel11824....1F,2020ApJ...889..142S,2021NatAs...5...94M,2021arXiv211113642T}.
Starting from 2018 July 4 (MJD 58303), the source underwent a rapid hard-to-soft transition. 
During the transition, an extremely powerful superluminal ejection was detected \citep{2020NatAs...4..697B} close in time to the appearance of the type-B QPO \citep{2020ApJ...891L..29H}.
After the transition, the source moved to the SS and stayed there for over 2 months before the final soft-to-hard transition (see \citealt{2020ApJ...889..142S} for the details of the outburst evolution).



\citet{2021MNRAS.506.2020D} studied the properties of the broadband noise in the PDS of \target\ with {\it NICER} data. They found that the broadband noise can be fitted with four Lorentzians and the spectra of these variability components are quite different in shape. At least two Comptonization regions with different temperatures and optical depths are required to fit both the variability spectra and the time-averaged spectra. \citet{2022MNRAS.511..536K} proposed a model based on the fluctuations propagation by considering that the hot inner flow is spectrally inhomogeneous, and the viscous time-scale is discontinuous between the disk and the hot flow. This model reconstructs the shape of the broadband noise below 10 keV in \target.
The large effective area of \hxmt\ at high energies enables us to perform detailed analysis on fast X-ray variability at energy bands above 100 keV \citep[e.g.,][]{2018ApJ...866..122H,2020SCPMA..6349503L,2021RAA....21...70L,2021cosp...43E1585H}. In \citet{2021NatAs...5...94M}, the authors reported the discovery of low-frequency QPOs above 200 keV in \target\ for the first time. In this work, we present a qualitative study of the evolution of the broadband noise and its energy dependence using \hxmt\ observations of \target. For the first time, we extend the study of the broadband noise up to 100--150 keV. 
In Section 2, we describe the observation and data reduction. The analysis and main results are presented in Section 3 and discussed in Section 4. In Section 5, we summarize our results.


\section{OBSERVATIONS AND DATA REDUCTION} \label{sec:style}

\hxmt\ is China's first X-ray astronomy satellite, launched on 2017 June 15 \citep{2020SCPMA..6349502Z}.
It carries three slat-collimated instruments: the High Energy X-ray telescope (HE: 20--250 keV, \citealt{2020SCPMA..6349503L}), the Medium Energy X-ray telescope (ME: 5--30 keV, \citealt{2020SCPMA..6349504C}), and the Low Energy X-ray telescope (LE: 1--15 keV, \citealt{2020SCPMA..6349505C}).

\begin{figure}
\centering
	\includegraphics[width=0.48\textwidth]{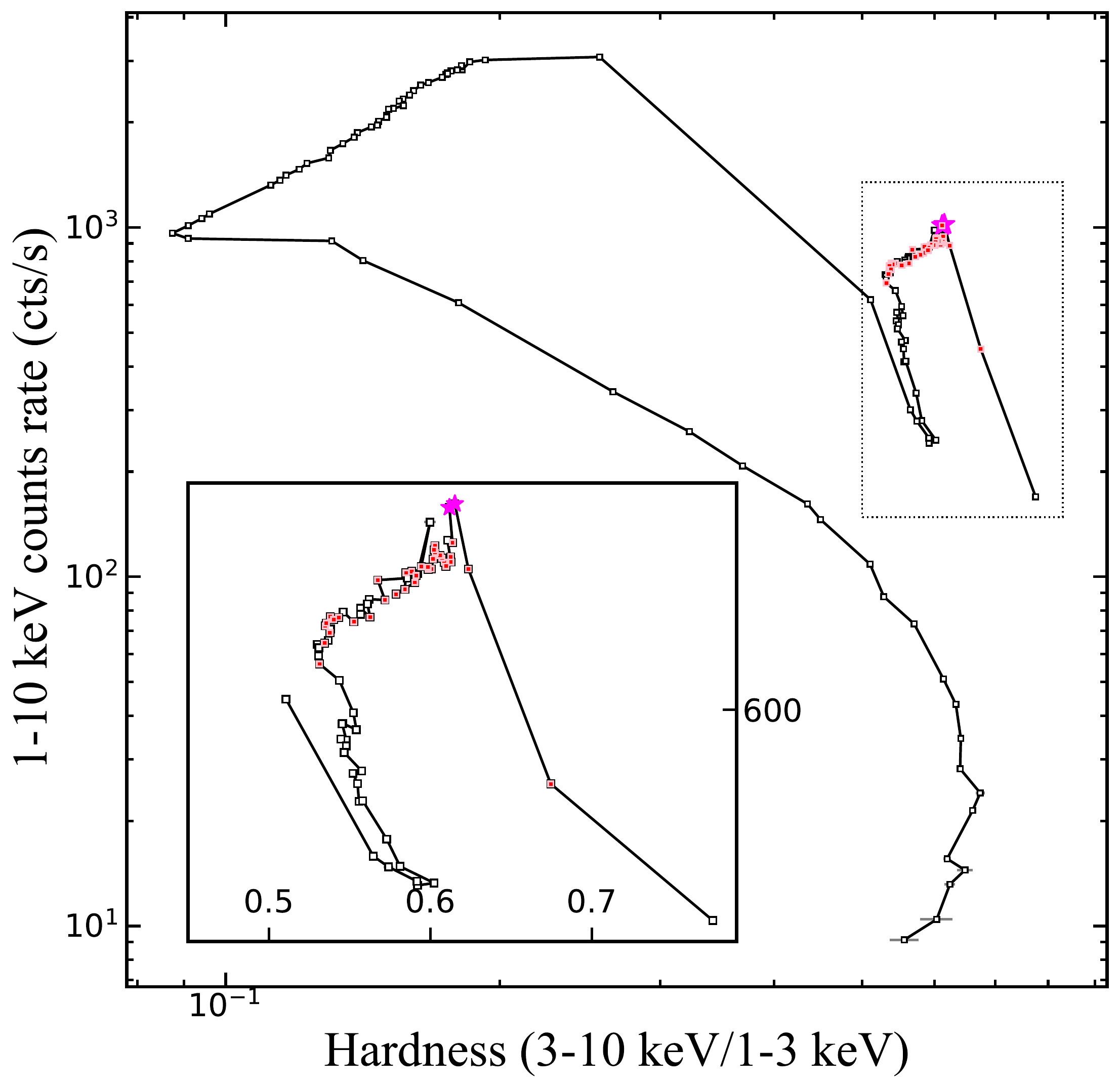}
    \caption{\hxmt\ hardness-intensity diagram (HID). Red points mark the observations used in this work to study the evolution of the different PDS components with hardness. Magenta points mark the observations we used to study the energy-dependent properties of these components.}
    \label{fig:HID}
\end{figure}

Four days after the discovery of \target\ , \hxmt\ started monitoring \target\ at a high cadence. During its initial full outburst between 2018 March and October, \hxmt\ accumulated a total exposure of 2560 ks. 
The LE (1--10 keV), ME (10--30 keV) and HE (30--150 keV) light curves of this outburst have been shown in \citet{2021NatAs...5...94M}. 
%
%
In Figure~\ref{fig:HID}, we show the HID of this outburst.
%
For our analysis, we only selected observations with the LE exposure time longer than 2500 s and the LE counts rate larger than 440 counts s$^{-1}$, where their PDS show a significant broadband noise and there are enough photons to do detailed timing analysis. Table \ref{tab:log} lists the log of the observations used in this work.

\startlongtable

  \begin{deluxetable*}{ccccc}
   \tablecaption{The log of the \hxmt\ observations used in this work. The hardness is the ratio of the count rate between the 1.0--3.0 keV and the 3.0--10.0 keV bands. \label{tab:log}}
\tablehead{
\colhead{ObsID} & \colhead{Start time (UTC)} & \colhead{Exposure (s)} & \colhead{Hardness} & \colhead{QPO}
}
\startdata
  P0114661002 & 2018/03/16 10:01:24 & 32717 & 0.67(1) & None\\
  P0114661003 & 2018/03/22 10:46:58 & 23245 & 0.62(1) & C\\
  P0114661004 & 2018/03/24 07:19:14 & 28893 & 0.61(1) & C\\
  P0114661005 & 2018/03/20 00:00:10 & 6343 & 0.62(1) & C\\
  P0114661006 & 2018/03/27 08:29:37 & 3325 & 0.61(1) & C\\
  P0114661008 & 2018/03/29 20:56:55 & 4649 & 0.61(1) & C\\
  P0114661009 & 2018/03/30 16:02:31 & 3728 & 0.61(1) & C\\
  P0114661010 & 2018/03/31 20:41:05 & 5356 & 0.61(1) & C\\
  P0114661011 & 2018/04/01 20:33:24 & 26322 & 0.61(1) & C\\
  P0114661012 & 2018/04/03 23:29:38 & 8122 & 0.61(1) & C\\
  P0114661013 & 2018/04/05 20:04:33 & 8490 & 0.61(1) & C\\
  P0114661014 & 2018/04/06 15:11:06 & 12809 & 0.60(1) & C\\
  P0114661015 & 2018/04/08 13:22:00 & 8151 & 0.60(1) & C\\
  P0114661016 & 2018/04/09 10:04:01 & 5744 & 0.60(1) & C\\
  P0114661017 & 2018/04/10 14:43:27 & 8337 & 0.60(1) & C\\
  P0114661018 & 2018/04/11 20:58:02 & 7313 & 0.60(1) & C\\
  P0114661019 & 2018/04/12 16:03:53 & 9654 & 0.60(1) & C\\
  P0114661020 & 2018/04/14 11:01:36 & 7600 & 0.60(1) & C\\
  P0114661021 & 2018/04/15 01:20:44 & 2992 & 0.59(1) & C\\
  P0114661024 & 2018/04/18 15:14:50 & 7265 & 0.59(1) & C\\
  P0114661025 & 2018/04/19 15:06:26 & 7360 & 0.59(1) & C\\
  P0114661026 & 2018/04/20 14:58:02 & 6492 & 0.59(1) & C\\
  P0114661027 & 2018/04/23 19:19:13 & 2812 & 0.58(1) & C\\
  P0114661028 & 2018/04/25 14:16:17 & 4639 & 0.58(1) & C\\
  P0114661029 & 2018/04/27 09:13:33 & 4215 & 0.58(1) & C\\
  P0114661031 & 2018/04/22 19:27:34 & 4025 & 0.58(1) & C\\
  P0114661032 & 2018/04/28 13:51:51 & 4029 & 0.57(1) & C\\
  P0114661035 & 2018/05/02 10:09:17 & 3617 & 0.56(1) & C\\
  P0114661038 & 2018/05/05 09:46:03 & 2817 & 0.56(1) & C\\
  P0114661040 & 2018/05/07 11:05:59 & 2719 & 0.55(1) & C\\
  P0114661041 & 2018/05/08 10:58:08 & 5513 & 0.54(1) & C\\
  P0114661042 & 2018/05/09 14:01:08 & 2632 & 0.54(1) & C\\
  P0114661043 & 2018/05/10 13:53:07 & 2633 & 0.54(1) & C\\
  P0114661044 & 2018/05/12 04:04:08 & 3950 & 0.54(1) & C\\
  P0114661045 & 2018/05/13 00:45:01 & 2729 & 0.54(1) & C\\
  P0114661048 & 2018/05/16 13:03:38 & 3388 & 0.54(1) & C\\
  P0114661052 & 2018/05/20 15:41:03 & 15578 & 0.53(1) & C\\
\enddata

\end{deluxetable*}

The data are extracted from all three instruments using the \hxmt\ Data Analysis software (HXMTDAS) v2.04 \footnote{The data analysis software is available from http://hxmten.ihep.ac.cn/software.jhtml.}, and filtered with the following criteria:  
(1) pointing offset angle less than $0.04\degree$;  
(2) Earth elevation angle larger than $10\degree$;  
(3) the vaule of the geomagnetic cutoff rigidity larger than 8 GV;  
(4) at least 300 s before and after the South Atlantic Anomaly passage.
To avoid possible contamination from the bright Earth and nearby sources, we only use data from the small field of view \citep{2018ApJ...864L..30C}.

\section{ANALYSIS AND RESULTS} \label{sec:floats}

\begin{figure}
\epsscale{1.0}
\centering
	\includegraphics[width=0.48\textwidth]{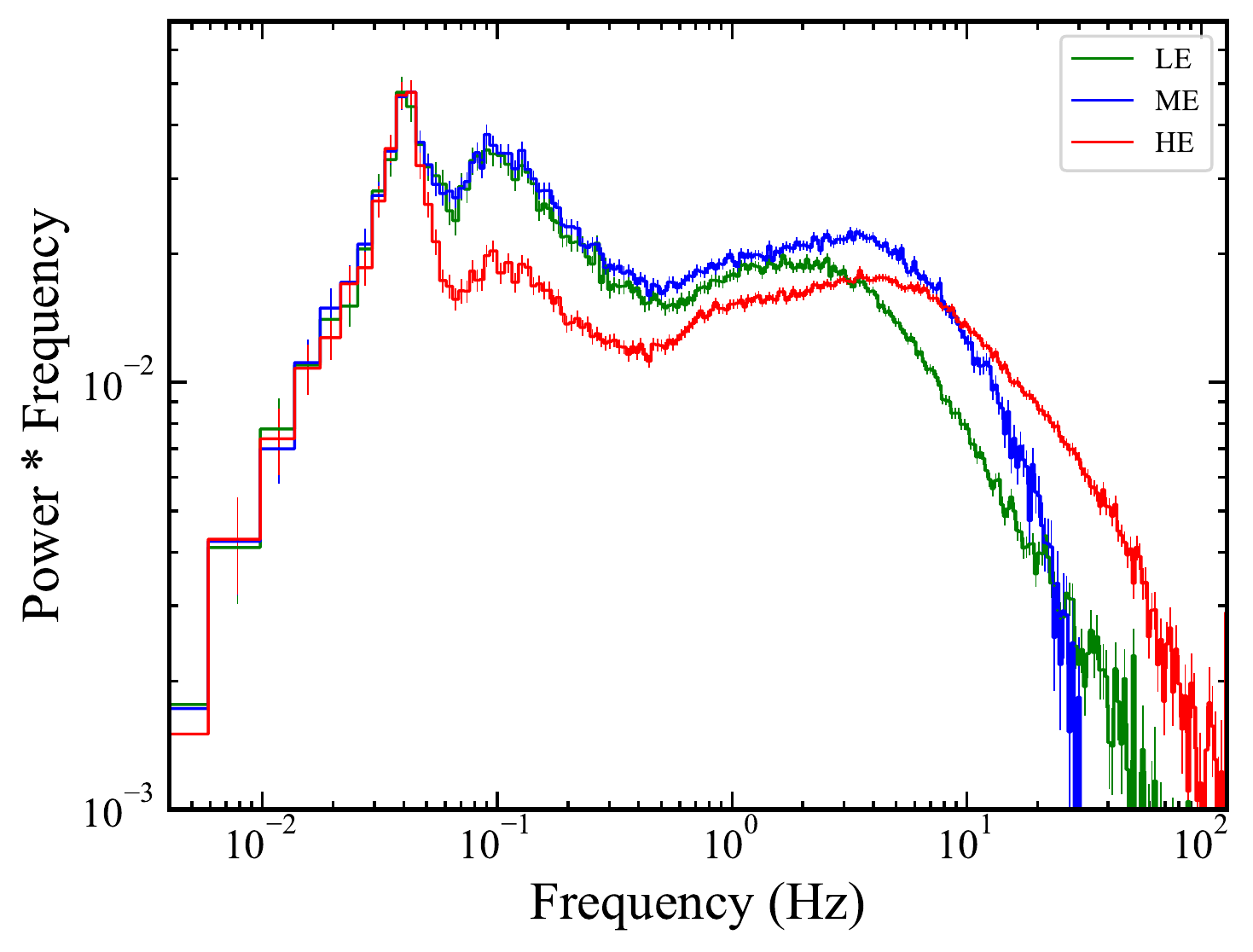}
    \caption{Representative PDS for LE (1--10 keV), ME (10--30 keV) and HE (30--150 keV), respectively. The PDS are calculated using the data of ObsIDs P0114661003 and P0114661004. The PDS of the two observations have a similar shape and consistent fractional rms of different energy bands. The PDS of ME and HE are separately multiplied by a factor of 1.4 and 2.1 to keep the QPO aligned between energy bands.}
    \label{LEMEHE_PDS}
\end{figure}

\begin{figure}
\epsscale{1.0}
\centering
	\includegraphics[width=0.48\textwidth]{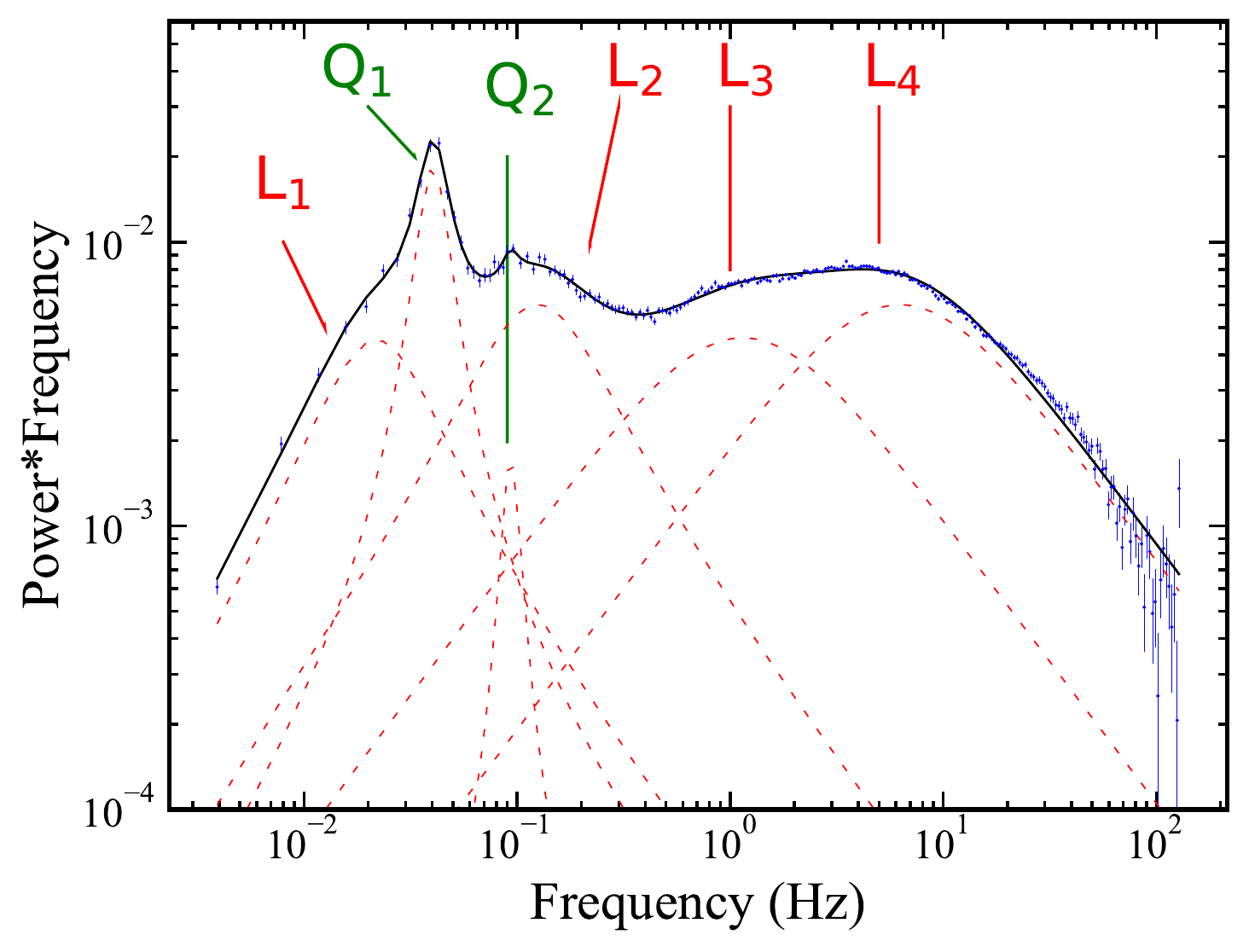}
    \caption{A representative PDS of HE (30-150 keV) plotted along with the best multi-Lorentzian fit (red). The PDS are calculated using the data of ObsIDs P0114661003 and P0114661004, and fitted with a multiple-Lorentzian model. $Q_1$ and $Q_2$ represent the QPO and its second harmonic, respectively. $L_{1}$, $L_{2}$, $L_{3}$ and $L_{4}$ represent the four broadband noise components on different timescales.}
    \label{HE50_60}
\end{figure}

To study the fast X-ray variability, we produce PDS from different energy bands for each observation we used. We use 128-s long intervals and 1/256-s time resolution, corresponding to a Nyquist frequency of 128 Hz. The PDS is then applied to Miyamoto normalization, namely normalized to fractional rms \citep{1991ApJ...383..784M}. 
In Figure~\ref{LEMEHE_PDS}, we show representative PDS for LE (1--10 keV), ME (10--30 keV) and HE (30--150 keV), respectively. The PDS of ME and HE are separately multiplied by a factor of 1.4 and 2.1 to keep the QPO aligned between energy bands. The PDS we show here are extracted from two relatively long observations (ObsIDs P0114661003 and P0114661004) with a similar PDS shape.
It is apparent from this figure that, although the shape of the noise component below the QPO frequency has not changed too much, the shape of the noise component above the QPO frequency changes significantly between energies.
%
In order to further study the properties of the QPO and the broadband noise, we fit the PDS with a multiple-Lorentzian model \citep{2002ApJ...572..392B}. 
%
In Figure~\ref{HE50_60}, we show a representative PDS of the HE (30--150 keV) band with its best fit.
The QPO and its second harmonic are fitted with one Lorentzian each ($Q_{1}$ and $Q_{2}$). 
The broadband noise shows a two-humped shape, and can be well-fitted with four Lorentzians, i.e., a very low-frequency noise ($L_{1}$), a low-frequency noise ($L_{2}$), and two high-frequency noise components ($L_{3}$ and $L_{4}$).
After the fitting process, we calculate the characteristic frequency and the fractional rms amplitude for each component. 
The characteristic frequency, $\nu_{\rm max}$, is defined as $\nu_{\rm max} = \sqrt{\nu_{0}^2 +  (\sigma/2)^2}$, where $\nu_{0}$ is the centroid frequency and $\sigma$ is the full width at half maximum (FWHM) of the Lorentzian function \citep{1997A&A...322..857B}.

In order to study the energy-dependent properties of the different components, we extract PDS from 12 energy bands: LE (1--3 keV, 3--7 keV), ME (7--10 keV, 10--20 keV, 20--30 keV), HE (30--40 keV, 40--50 keV, 50--60 keV, 60--70 keV, 70--80 keV, 80--90 keV, 90--150 keV).
In this paper, we only show the results combined from ObsIDs P0114661003 and P0114661004. The energy-dependent properties of the other observations are similar. 

\subsection{Properties of $L_{1}$}

\begin{figure}
	\includegraphics[width=0.48\textwidth]{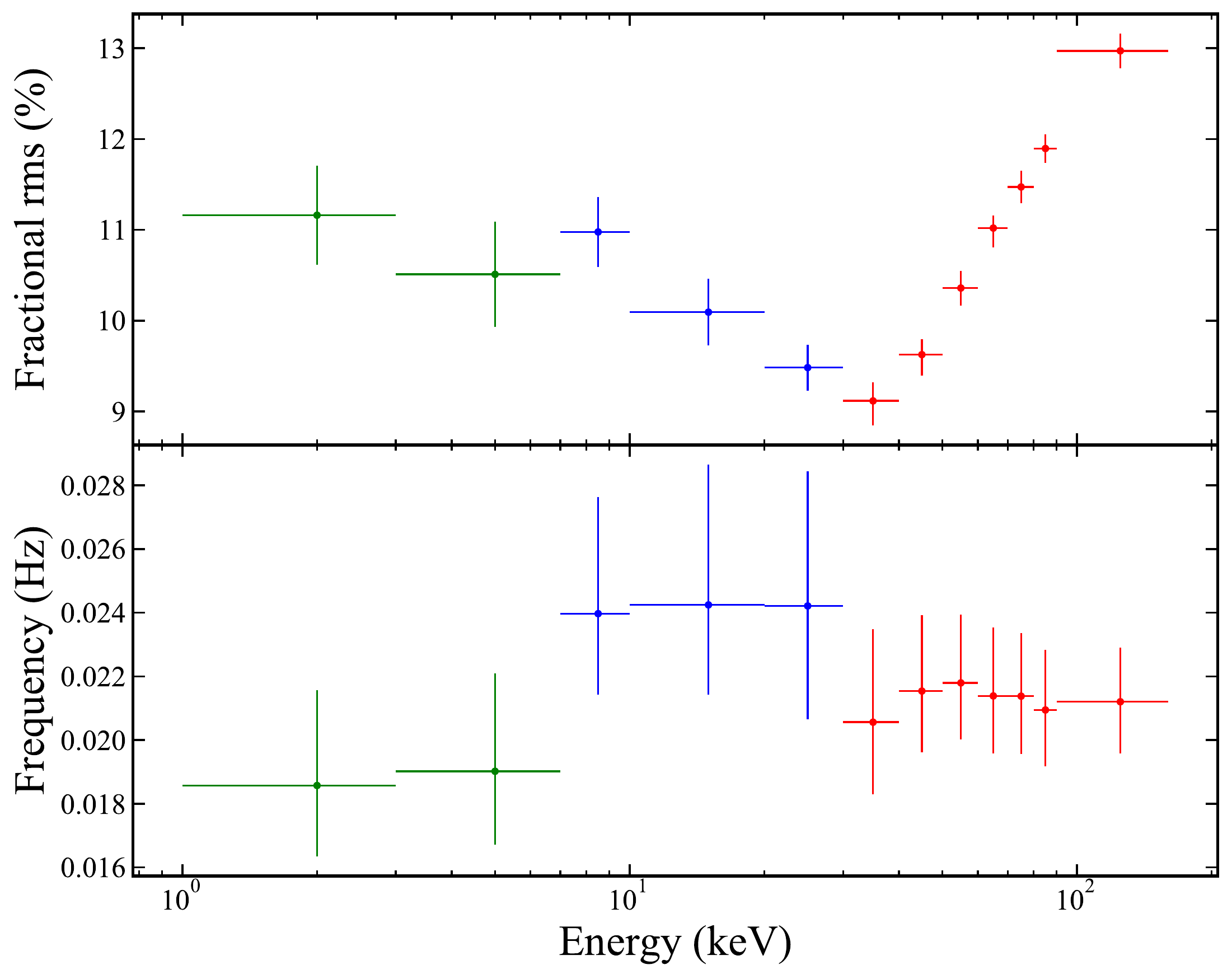}
    \caption{Energy dependence of the fractional rms and characteristic frequency of $L_{1}$. Green, blue and red points represent the LE, ME and HE data, respectively. The characteristic frequency does not show significant changes with energy. Therefore, we fixed the centroid frequency and FWHM of $L_{1}$ when calculating the energy dependence of the fractional rms. The data are extracted from ObsID P0114661003 and P0114661004.}
    \label{L1}
\end{figure}

 Figure~\ref{L1} shows the energy dependence of the fractional rms and characteristic frequency of $L_{1}$. The characteristic frequency of $L_{1}$ remains more or less constant in the 1--150 keV energy band. This is similar to what was found in \citet{2021NatAs...5...94M} for the QPO and its second harmonic. The evolution of the fractional rms of $L_{1}$ with energy is more complicated. Below $\sim$30 keV, the fractional rms shows a slight decreasing trend with energy; while above $\sim$30 keV, the fractional rms increases monotonously with energy.  
 
\begin{figure}
\epsscale{1.0}
\centering
	\includegraphics[width=0.48\textwidth]{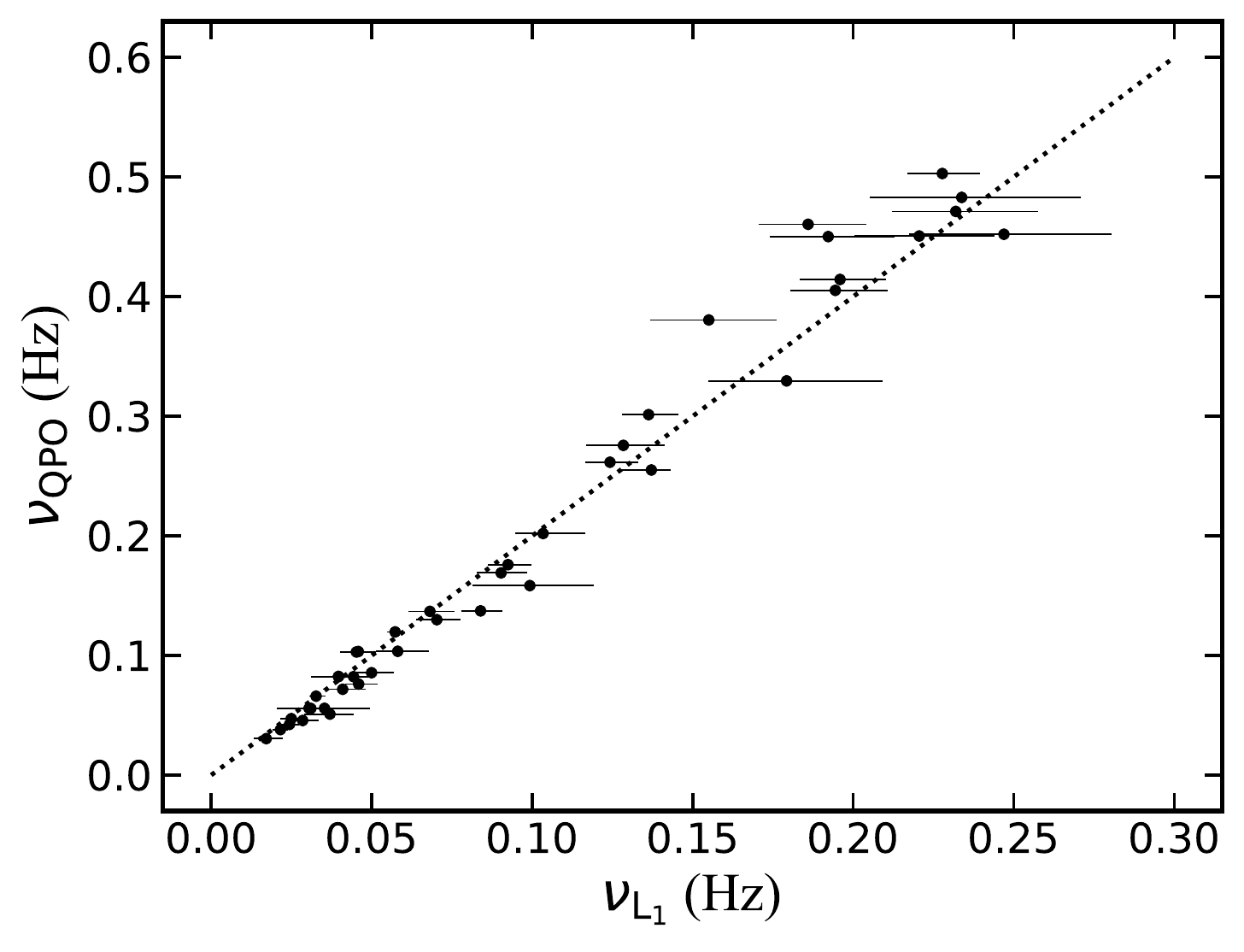}
    \caption{Characteristic frequency of the QPO as a function of the characteristic frequency of $L_{1}$. The dotted line represents the correlation $\nu_{\rm QPO} = 2\nu_{L_{1}}$. The characteristic frequencies are measured by fitting the PDS of HE band. 
    }
    \label{wk}
\end{figure}

 %
 In Figure~\ref{wk} we show the characteristic frequency of the QPO as a function of the characteristic frequency of $L_{1}$. It can be seen that the data points follow the correlation $\nu_{\rm QPO} = 2\nu_{L_{1}}$, rather than the WK correlation found by \citet{1999ApJ...514..939W} and \citet{2017ApJ...841..122B} in other BHXRBs. In LHS, the correlation between characteristic frequencies of QPO and low frequency broadband noise can be fitted by a power-law function the so called WK correlation. This suggests that $L_{1}$ is more like a sub-harmonic of the QPO rather than broadband noise. We have tried to fit the low-frequency part of the PDS (below the QPO frequency) with two Lorentzian functions. However, adding another Lorentzian does not improve the fits a lot and this extra component is not statistically needed. Note that, the quality factor for $L_{1}$ is relatively low (< 1) compared with QPO $Q_{1}$ and harmonic $Q_{2}$.
 
\begin{figure}
\epsscale{1.0}
\centering
	\includegraphics[width=0.48\textwidth]{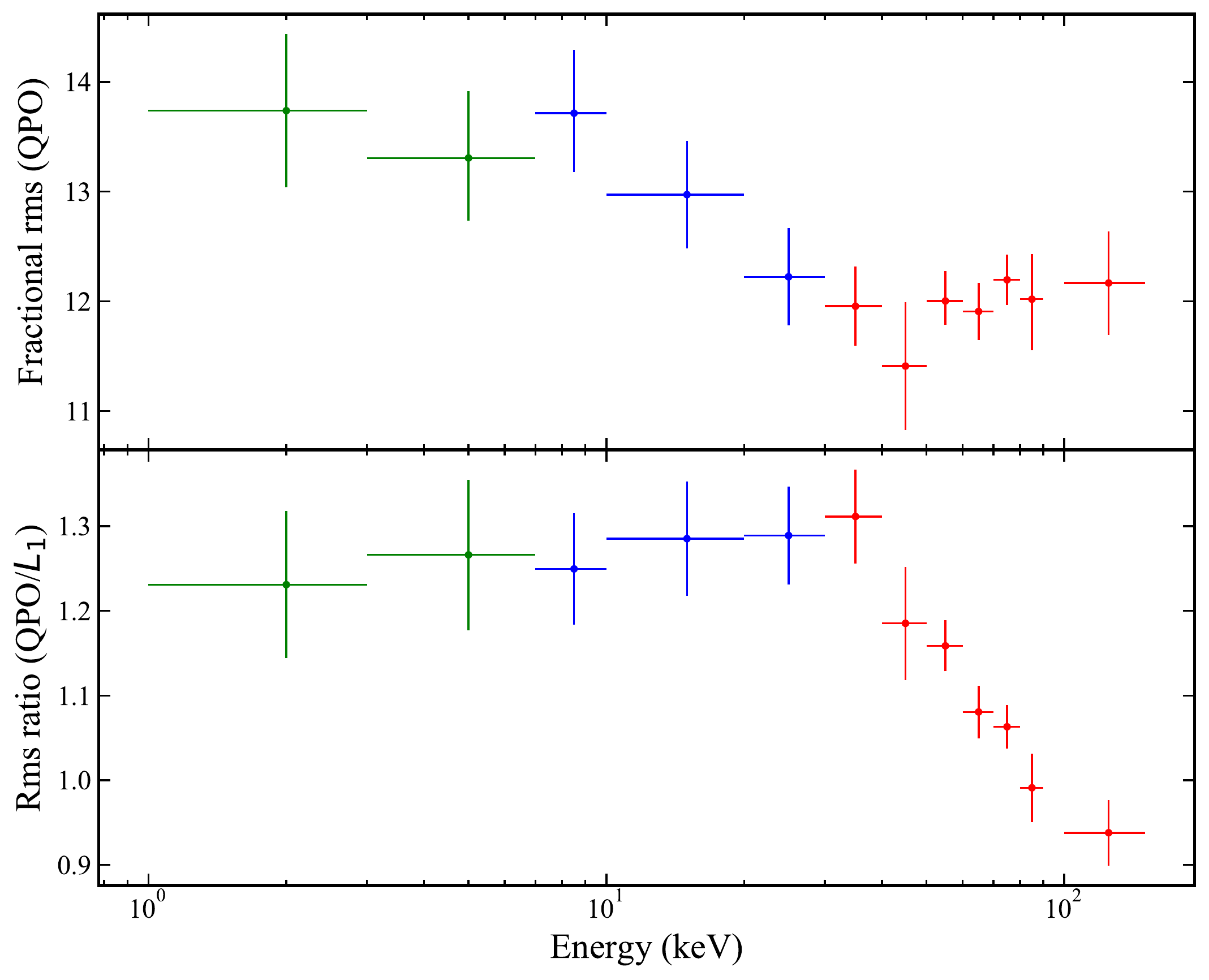}
    \caption{The fractional rms of QPO (top panel) and the ratio of fractional rms between QPO and $L_1$ (bottom panel) versus energy band. The data are extracted from ObsID P0114661003 and P0114661004.}
    \label{ratio_L1_qpo}
\end{figure}

 In order to compare the fractional rms relation between $L_{1}$ and $Q_{1}$, we show the rms(E) of QPO and the ratio of rms between $Q_{1}$ and $L_{1}$ in Figure~\ref{ratio_L1_qpo}. The ratio keeps almost constant at 1.2--1.3 when energy is lower than 30 keV. However, when energy is higher than 30 keV, the ratio starts to decline sharply from 1.3 to 0.87.

\subsection{Properties of $L_{2}$, $L_{3}$ and $L_{4}$}
\subsubsection{Evolution with spectral hardness}

\begin{figure}
\epsscale{1.0}
\centering
	\includegraphics[width=0.48\textwidth]{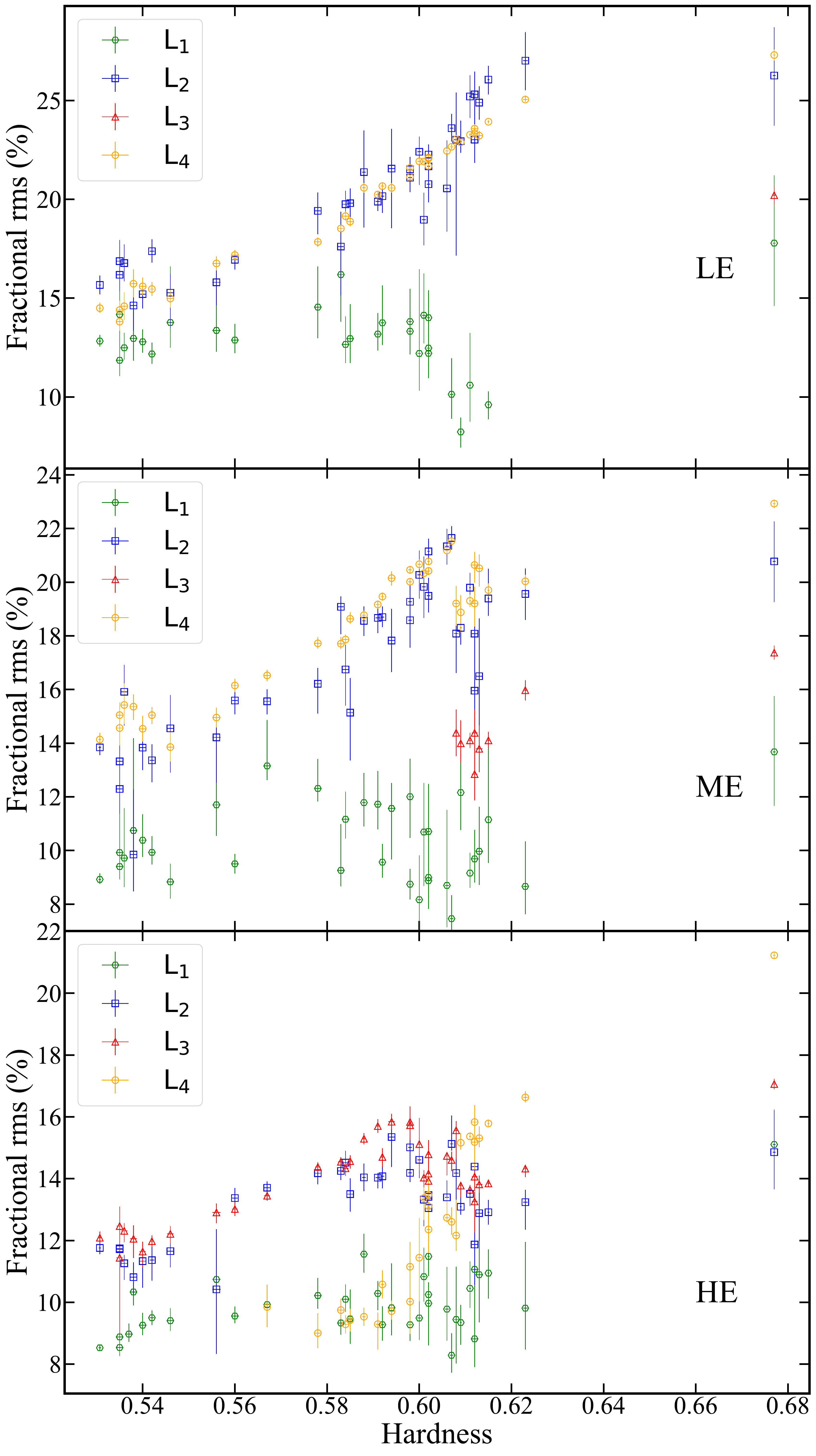}
    \caption{Evolution of the fractional rms of $L_{1}$, $L_{2}$, $L_{3}$ and $L_{4}$ with hardness. From top to bottom, the fractional rms are calculated in the LE (1--10 keV), ME (10--30 keV) and HE (30--150) keV bands, respectively.}
    \label{h-r}
\end{figure}

Hereafter we mainly focus on the noise components above the QPO frequency. Figure~\ref{h-r} shows the evolution of the fractional rms of $L_{2}$, $L_{3}$ and $L_{4}$ with hardness ratio. From top to bottom, the rms are calculated in the LE (1--10 keV), ME (10--30 keV) and HE (30--150) keV bands, respectively. In all panels, the hardness ratio is defined as the ratio of the count rate between the 1--3 keV and the 3--10 keV bands.
%
%
In the LE and ME bands, the fractional rms of all the three components generally decrease as the spectrum softens. However, in the HE band, the evolution of the fractional rms is more complicated. In the hardness range $\sim$0.54--0.60, the fractional rms of $L_{2}$ and $L_{3}$ increase with hardness; while the fractional rms of $L_{4}$ remains almost constant. In the hardness range $\sim$0.60--0.62, the fractional rms of $L_{2}$ and $L_{3}$ instead tend to decrease; whereas the fractional rms of $L_{4}$ starts increasing sharply. In the hardness range $>0.62$, we do not have a good monitoring coverage. Overall, it seems that the fractional rms of all the three components show an increasing trend with hardness.

\begin{figure}
\epsscale{1.0}
\centering
	\includegraphics[width=0.48\textwidth]{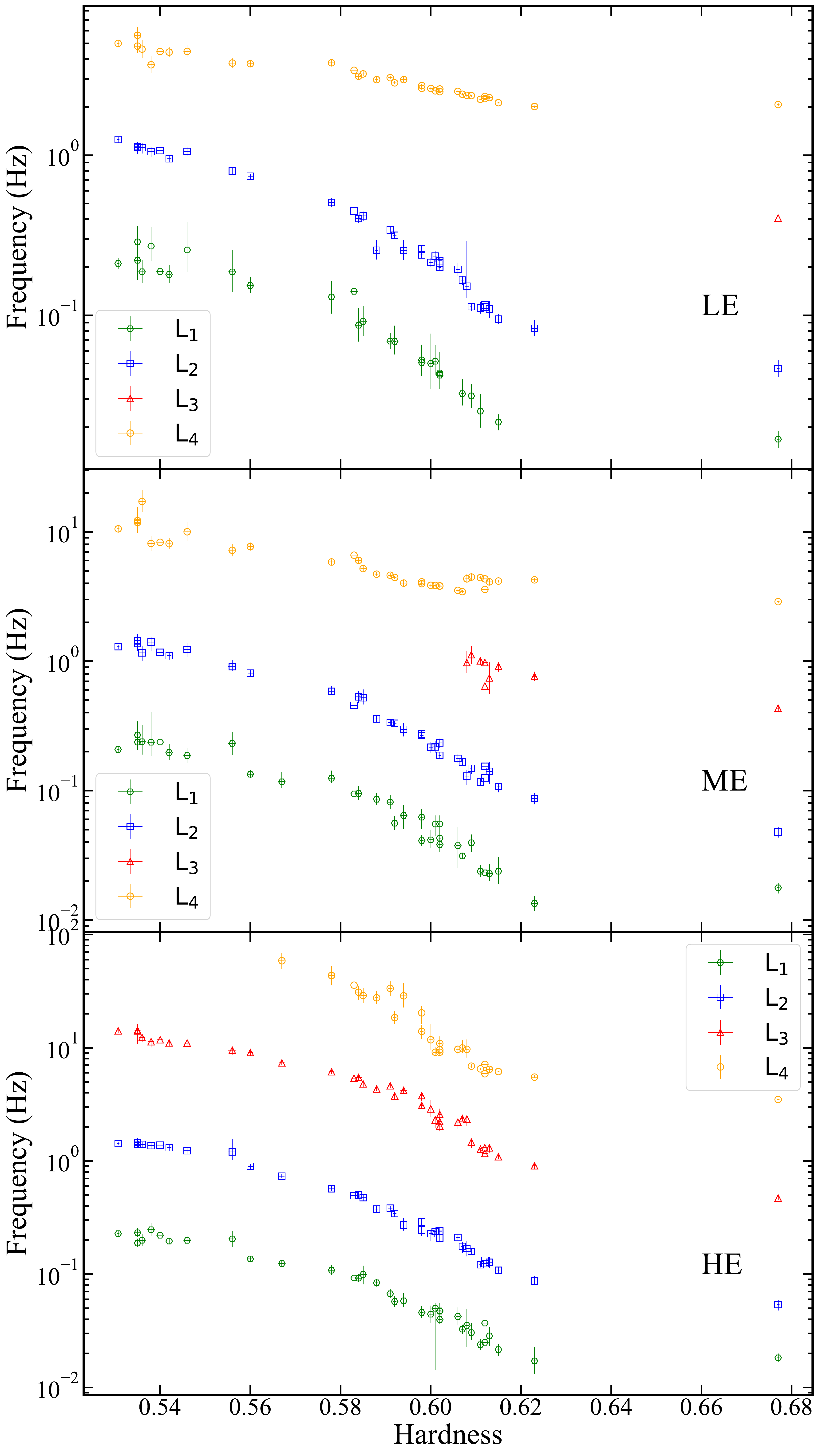}
    \caption{Evolution of the characteristic frequencies of $L_{1}$, $L_{2}$, $L_{3}$ and $L_{4}$ with hardness. }
    \label{h-f}
\end{figure}

Figure~\ref{h-f} shows the corresponding evolution of the characteristic frequencies of the three components with spectral hardness. It can be seen that the characteristic frequencies of all the three components generally increase as the spectra softens. At hardness > $\sim$0.62, the increasing trend is seems to be flatter than that at hardness < $\sim$0.62.

It is worth noting that $L_{3}$ is not always present in all cases. In the PDS of LE, $L_{3}$ is only seen in the observation (ObsID P0114661002) where the spectrum is the hardest. In the PDS of ME, it only appears when the hardness ratio is larger than $\sim$0.61. While in the HE band, we can see this component in all observations.

Except for the evolution trend with hardness ratio for $L_2$, $L_3$ and $L_4$, we also plot the evolution of the fractional rms, characteristic frequency for $L_1$ component in Figures~\ref{h-r} and \ref{h-f}. As we can see from Figure~\ref{h-r}, the fractional rms of $L_1$ shows no obvious evolutionary trend with hardness ratio which is totally different from other broadband noise components. In contrary, from Figure~\ref{h-f}, the evident decreasing trend of $L_1$ implicates the physical relation between radiation region and spectral evolution.

\subsubsection{Energy-dependent properties}

\begin{figure*}
\epsscale{1.0}
\centering
   \includegraphics[width=\textwidth]{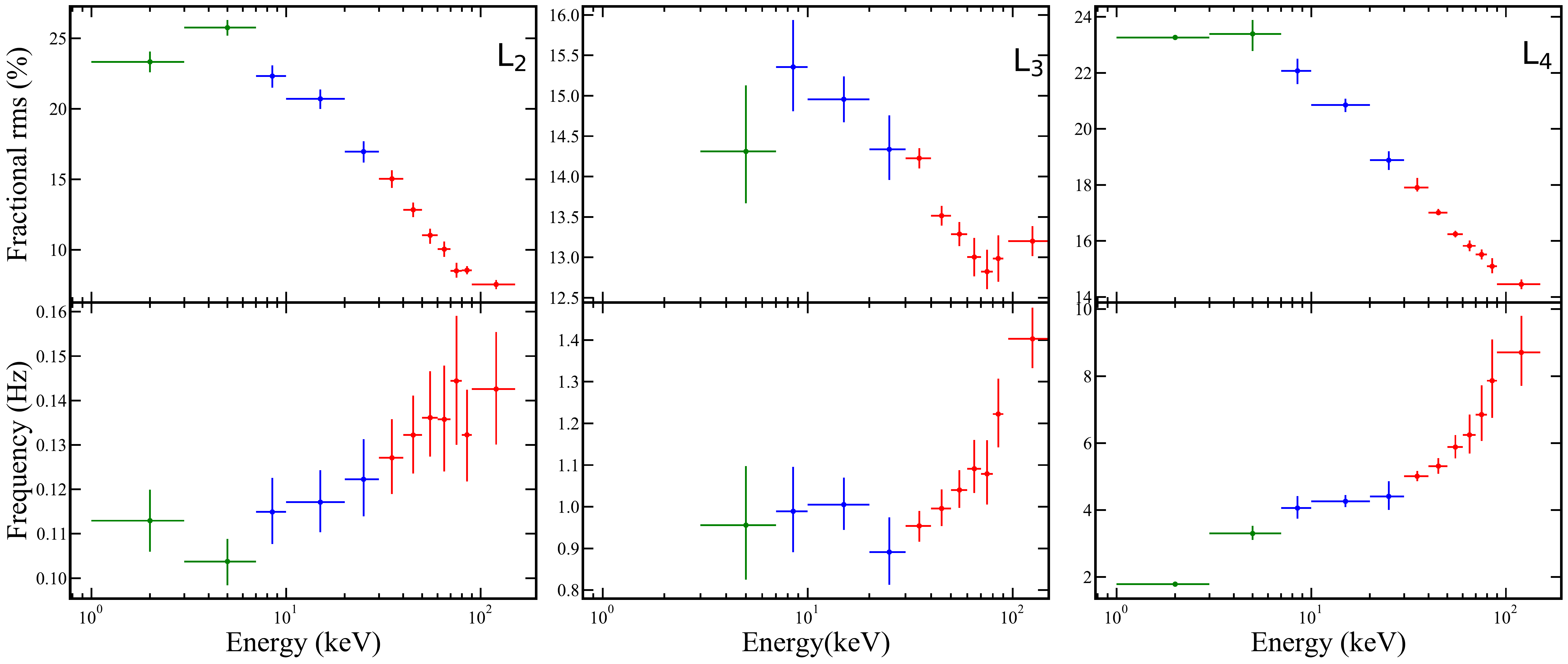}
   \caption{Energy dependence of the fractional rms and characteristic frequency for $L_{2}$ (left),$L_{3}$ (middle) and $L_{4}$ (right). Green, blue and red points represent the LE, ME and HE data, respectively.}
   \label{combine}
\end{figure*}

Figure~\ref{combine} shows the energy dependence of the fractional rms and characteristic frequency of $L_{2}$ (left), $L_{3}$ (middle) and $L_{4}$ (right). As we can see from Figure ~\ref{combine} right panel, fractional rms decreases from 23\% to 15\% with increasing energy while characteristic frequency always increases with energy. In the left panel ($L_{2}$), the fractional rms shows the same trend but the characteristic frequency first keeps almost constant below 20--30 keV, whereas above 20--30 keV, the frequency increases rapidly with energy. This case is also true in middle panel ($L_{3}$), although with a large uncertainty below 20 keV. It is worth noting that the characteristic frequency of $L_{4}$ above 90 keV is almost five times the frequency in 1--10 keV (from $\sim$ 2 Hz to $\sim$ 10 Hz). Similar results in LE band can be found in \citet{2022MNRAS.511..536K}. They used {\it NICER} data to study the relationship between asymmetric Lorentzian function P1, P2 and energy which actually reflects the evolution of characteristic frequency with energy for $L_{2}$, $L_{3}$ and $L_{4}$ components (P1 corresponds $L_{2}$; P2 corresponds $L_{3}$ and $L_{4}$)\footnote{The ObsID we choose and \citet{2022MNRAS.511..536K} choose are no more than one day apart, so we can easily match P1, P2 with $L_{2}$, $L_{3}$ and $L_{4}$. P1 corresponds to $L_{2}$, P2 corresponds to $L_{3}$ and $L_{4}$.}. In order to fit the PDS phenomenally, \citet{2022MNRAS.511..536K} used two asymmetric Lorentzian functions. Nevertheless, considering the Lorentzian function that fluctuation propagation predicts, we decide to use three standard Lorentzian functions to fit PDS.

\begin{figure}
	\includegraphics[width=0.48\textwidth]{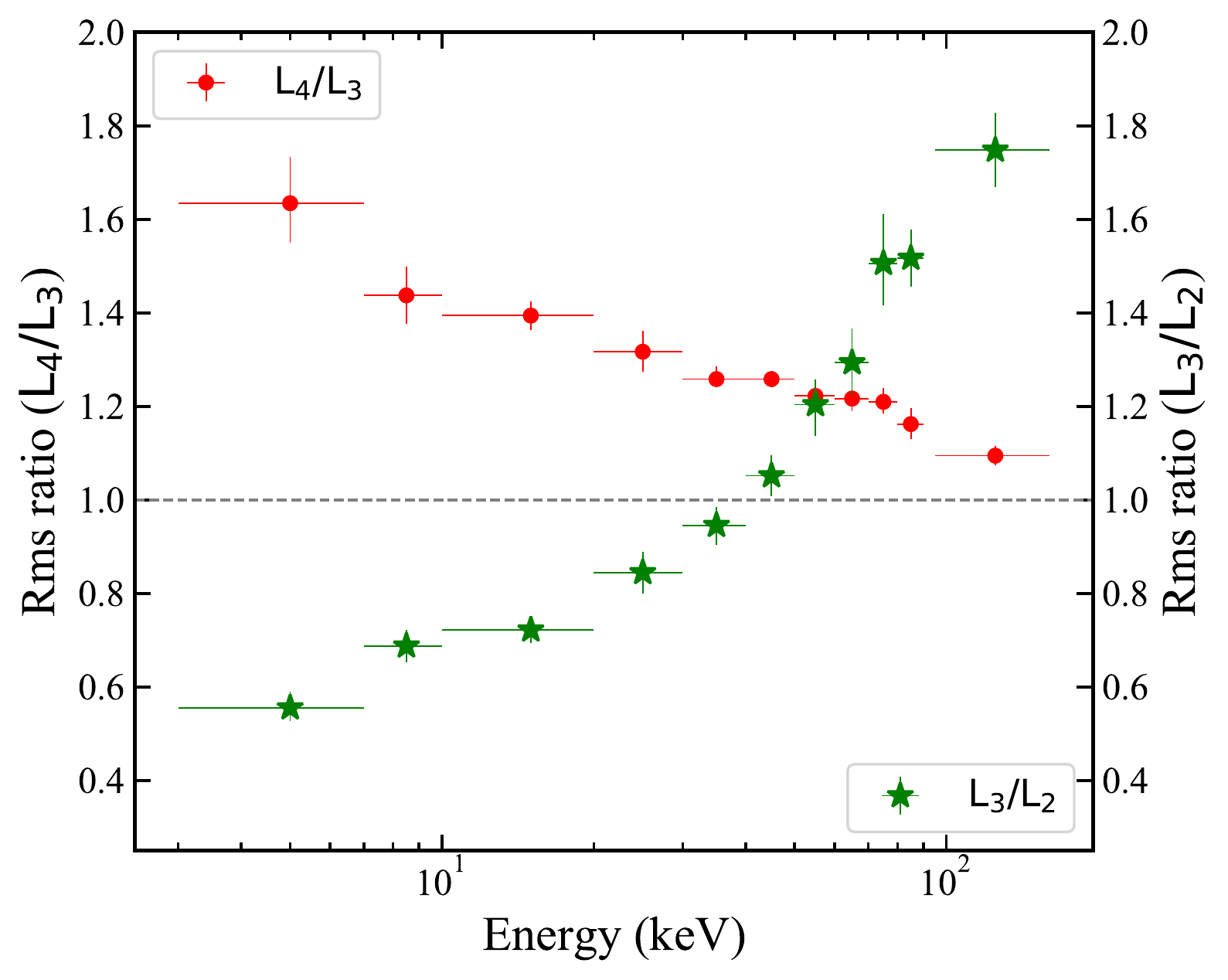}
    \caption{The ratio of fractional rms between $L_{2}$ and $L_{3}$ (green points), $L_{3}$ and $L_{4}$ (red points) versus energy band. The data points are extracted from ObsID P0114661003 and P0114661004.}
    \label{ratio}
\end{figure}

In order to compare the relation between fractional rms of $L_{2}$, $L_{3}$ and $L_{4}$, Figure~\ref{ratio} shows the energy dependence of fractional rms ratio for three different Lorentzian components. As we can see from Figure~ \ref{ratio}, with increasing energy, when energy is below 20--30 keV, the ratio between $L_{3}$ and $L_{2}$ changes slightly and is less than 1. When energy is higher than 20--30 keV, the rms ratio starts to increase robustly  to 1.8. As for $L_{4}$ and $L_{3}$, unlike $L_{2}$, there is a totally different trend. The rms ratio between $L_{4}$ and $L_{3}$ decreases from 1.6 to 1.1 with increasing energy. Although the downward trend is opposite to the upward trend for $L_{3}$/$L_{2}$, but the rms ratio is still larger than 1 which means $L_{4}$ is more variable than $L_{3}$ at high energy band.

\subsection{Phase lag spectra}
Figure~\ref{pdslag} shows a representative power spectrum and corresponding phase lag spectrum.  At higher frequency above QPO frequency, hard lag feature is prominent. In logarithmic coordinates, the shape of hard lag is similar to a normal distribution with large normalization plus another normal function with small normalization at low frequency (see Figure~\ref{pdslag}). We will only focus on the high frequency (higher than QPO frequency) part corresponding to the broadband noise components. Then we use two log-normal functions to fit the hard lag part to get the peak value frequency of two log-normal functions (hereafter called $\nu_{\rm g1}$ and $\nu_{\rm g2}$). Meanwhile, we can approximately see that $\nu_{\rm g1} \simeq \nu_{\rm L3}$ and $\nu_{\rm g2} \simeq \nu_{\rm L2}$.  Then, we produce the phase lag spectrum for different energy band, with reference to the 1--10 keV band \citep{1999ApJ...510..874N,2015MNRAS.449.4027A,2017ApJ...845..143Z}. The phase lag spectrum are similar between different energy bands as Figure~\ref{pdslag} shows and has a similar shape as the phase lag spectra given by \citet{2021NatAs...5...94M}. As Figure~\ref{lagm} clearly shows, as the photon energy increases, the phase lag peak value for two normal functions also increases with energy band. 

\section{DISCUSSION} 
In this paper, we have investigated the evolution of the broadband noise in the PDS of \target\ using observations from \hxmt. We uncover the possible sub-harmonic component, $L_{1}$, and find the different energy dependence of fractional rms, characteristic frequency. We extend the study of the energy-dependent broadband noise up to 100--150 keV for the first time. It is found that the fractional rms of $L_{2}$, $L_{3}$ and $L_{4}$ generally increase with hardness in LE/ME/HE band, whereas characteristic frequency decreases with hardness. As for energy dependence, the rms of all three components decreases with energy. In particular, the characteristic frequencies of $L_{2}$ and $L_{3}$ remain unchanged below 20--30 keV then increase to 150 keV. The characteristic frequency of $L_{4}$ always increases with energy from 1--150 keV.

\subsection{Properties of $L_1$}
As shown in Section 3.1,  we suggest that the peak of $L_1$ observed at half the QPO frequency. This implicates that the $L_{1}$ component is the sub-harmonic of QPO with low Q value. In the jet precessing model \citet{2021NatAs...5...94M} proposed, the QPO signal is believed to originate from the precession of a small-scale jet to explain the LFQPO's high energy, soft lag and the maximum value, and energy-related behaviours for frequency, fractional rms and phase lag. As a result, considering the tight relation in Figure~\ref{wk}, we suggest that the same origin from small-scale jet for $L_{1}$. However, we cannot provide a physical model to interpret QPO and harmonics components at the same time \citep{2010ApJ...714.1065R,2012MNRAS.423..694R}. Considering the poor understanding for harmonics and our study mainly focusing on the broadband noise component, we will mainly discuss the $L_2$, $L_3$ and $L_4$. 

\subsection{Evolution of the fractional rms and characteristic frequency with hardness}

As shown in Figure~\ref{h-f}, the characteristic frequencies of all the high-frequency components ($L_{2}$, $L_{3}$ and $L_{4}$) increase with spectral softening. This is consistent with the evolution trend commonly found in other BHXBs  \citep{1999ApJ...520..262P,2015ApJ...806...90B,2015ApJ...805..139Z}. The trend of the characteristic frequency to increase with spectral softening can be explained naturally under the truncated disk/corona geometry \citep{1997ApJ...489..865E,2007A&ARv..15....1D}.
In the truncated disk model, the outer part of the accretion flow forms a geometrically thin, optically thick accretion disk truncated at a larger radius. The inner part of the flow is a hot, geometrically thick and optically thin configuration. The reduction in hardness ratio generally means a higher accretion rate, thus corona and disk should also move closer to black hole \citep{2005A&A...440..207B,2010LNP...794...53B}. In the fluctuation propagating model \citet{1997MNRAS.292..679L,2001MNRAS.327..799K} proposed, characteristic frequency is related to the outer radius of hot flow, and as the accretion rate increases, the characteristic frequency increases while outer radius decreases. Hence the viscous frequency of each component also increases with decreasing hardness.  
%

In view of fractional rms, with increasing hardness, the fractional rms also increases in $L_{2}$, $L_{3}$ and $L_{4}$. Based on truncated disk/corona model, corona shows more variability than standard accretion disk \citep{2006MNRAS.370..405S,2013MNRAS.431.1987A}. Consequently, when hardness ratio increases, more corona components contribute a higher fractional rms value.

In summary, the different components show similar evolutionary trends for fractional rms, characteristic frequency with hardness ratio in LE/ME/HE energy bands. The evolutionary trend can be explained under the truncated disk/corona model \citep{1997ApJ...489..865E,2007A&ARv..15....1D}. Actually, there are also still debates on the truncation of accretion disk in MAXI J1820+070. \citet{2019MNRAS.490.1350B} discovered a steady inner accretion disc measured by relativistic reflection. \citet{2019Natur.565..198K} found that the reverberation time lags between the continuum-emitting corona and the irradiated accretion disk are much shorter than previously seen in truncated accretion disk, and the timescale of the reverberation lags shortens by an order of magnitude over a period of weeks, whereas the shape of the Fe K$\alpha$ emission line remains remarkably constant. The similar results are also obtained from spectral analysis. Meanwhile, there are some other studies that support the truncated accretion disk argument from either spectral analysis or timing analysis \citep{2021A&A...654A..14D,2021A&A...656A..63M,2021ApJ...909L...9Z,2021ApJ...914L...5Z}. \citet{2021A&A...654A..14D} found that the frequency of thermal reverberation lags increases steadily, and, on the other hand, the temperature of the quasi-thermal component grows as the source softens, which can be explained in terms of a decrease in the disc inner radius. Moreover, \citet{2021A&A...654A..14D}  measured that the values of lag amplitude are a factor of 3 longer than those reported in \citet{2019Natur.565..198K}. The longer lags might not be easily reconciled with the conclusion of a disc extending close to the ISCO. \citet{2021ApJ...909L...9Z,2021ApJ...914L...5Z} confirm the optically-thick disk at least $> 10R_g$ from joint spectral analysis. To sum up, so far, all arguments in favor of the non-truncated disk model can be reasonably explained under the truncated disk model. Except for the methods mentioned above, in the present paper, we confirm the truncated accretion disk model in MAXI J1820+070 from the evolutionary trend of broadband noise components with a new perspective by means of the correspondence relation between break frequency and radiation region radius. Quantitatively, we can calculate the radiation region at different energy band of $L_2$ which represents the variable emission from the outermost region (see Figure~\ref{R-E}). We take a parameter set for standard $\alpha$-disk: $\alpha$=0.1, $ M_{\rm BH}=10\rm M_{\sun}$, scale height $H/r = 0.1$ to calculate the viscous frequency at certain radius \citep{2008bhad.book.....K}. As Figure~\ref{R-E} shows, the characteristic frequency of $L_2$ component shows energy-dependence: the emitting region spans from $\sim 34 R_g$ to $\sim 27 R_g$ corresponding to 1--150 keV photon energy. Like \citet{2021MNRAS.506.2020D} and \citet{2022MNRAS.511..536K} indicated, with frequency-resolved spectral analysis, the $L_2$ component is supposed to come from variable disc emission. However, making use of the ME and HE data from \hxmt, we actually detect the emission from high energy emission $> 100$ keV from the $L_2$ component which cannot be attributed to a simple standard accretion disc. Considering the change of radius plotted in Figure~\ref{R-E}, even though we can attribute the high energy emission to the propagation of fluctuation from disc to hot flow, then we should also expect a constant characteristic frequency for $L_2$ at high energy band according to fluctuation propagation. The characteristic frequency of $L_2$ remains unchanged below 20--30 keV as \citet{2022MNRAS.511..536K} found the constant peak frequency for P1. However, when the photon energy is greater than 20--30 keV, the radiation radius of $L_2$ starts to decrease to 27$R_g$. This phenomenon cannot be easily interpreted by fluctuation from disk propagating to hot flow. Therefore, we speculate that $L_2$ may be originated from a warm extended variable disk region. We therefore should consider more complicated accretion flow geometry where a standard accretion disc transits to a hot ADAF geometry.

\subsection{Energy dependence of broadband noise}

First, the fractional rms of $L_{2}$, $L_{3}$ and $L_{4}$ all generally show decreasing trend with energy. This phenomenon also exists for another black hole transient MAXI J1348-630 for broadband noise component in LHS \citep{2021cosp...43E1585H}. This phenomenon that fractional rms decreases with energy can be interpreted as variable input soft photon flux connected with Comptonization process \citep{2005MNRAS.363.1349G}. Generally speaking, this can be connected with an accretion disk located outer corona. Because of Magnetorotational Instability (MRI), the magnetic field will excite the fluctuation of accretion rate in accretion disk \citep{1991ApJ...376..223H,1998RvMP...70....1B,2005ASPC..330..185B,2008MNRAS.390...21B,2011ApJ...730...36D}. The fluctuation of mass accretion rate will cause variable emission. Then this mechanism will cause variable input soft photons flux. The variable seed photon is up-scattered in corona to higher energy radiation. Besides, by comparing the rms ratio between $L_{2}$, $L_{3}$ and $L_{4}$, we can find that with energy increasing, more and more variable high energy photon comes from $L_{3}$ than $L_{2}$. In other words, the radiation region that $L_{3}$ corresponds to should be hotter than $L_{2}$. As for $L_{4}$ and $L_{3}$, the decreasing trend can be interpreted by different variable seed photon flux in Comptonization process. If we consider a radially-extended corona, $L_{3}$ comes from outer region whereas $L_{4}$ comes from inner region, the inner region contributes more radiation than outer region at 1--150 keV. In other words, the rms ratio between $L_{4}$ and $L_{3}$ should always be higher than 1. Meanwhile, we should note that the inner region receives more variable soft photons than outer region (inner region not only receives soft photons from disk but also from the outer region, the flux from the outer region should be more variable than the flux from standard accretion disk, this phenomenon has been found in \citet{2021MNRAS.506.2020D}, most of the disc photons upscattered in the outer Comptonization region (Zone II) are used as seed photons for the inner Comptonization region), therefore, the decreasing slope of fractional rms with energy for $L_{4}$  should be greater than $L_{3}$ which means a downward trend for $L_{4}$/$L_{3}$ rms ratio with energy.  

Then, the energy dependence of characteristic frequency is comparatively complicated. For $L_{2}$ and $L_{3}$ components, the characteristic frequency keeps almost constant below 20--30 keV, then it increases with energy up to 90--150 keV. For $L_{4}$ component, the characteristic frequency always increases with energy from 1--150 keV. When energy is below 20--30 keV, the constant characteristic frequency for $L_{2}$ and $L_{3}$ may reflect the relatively uniform radiation area in the outer region. However, when energy is above 30 keV, the increasing characteristic frequency of $L_{2}$ and $L_{3}$ may reflect the increasing optical depth in hot flow from outer to inner region (This will be discussed in Section 4.4).

In summary, the energy dependence of fractional rms, characteristic frequency indicates that a complicated stratified accretion flow consisting of multiple coronae is need.

\subsection{Phase lag spectra}
\begin{figure}
\epsscale{1.0}
\centering
	\includegraphics[width=0.48\textwidth]{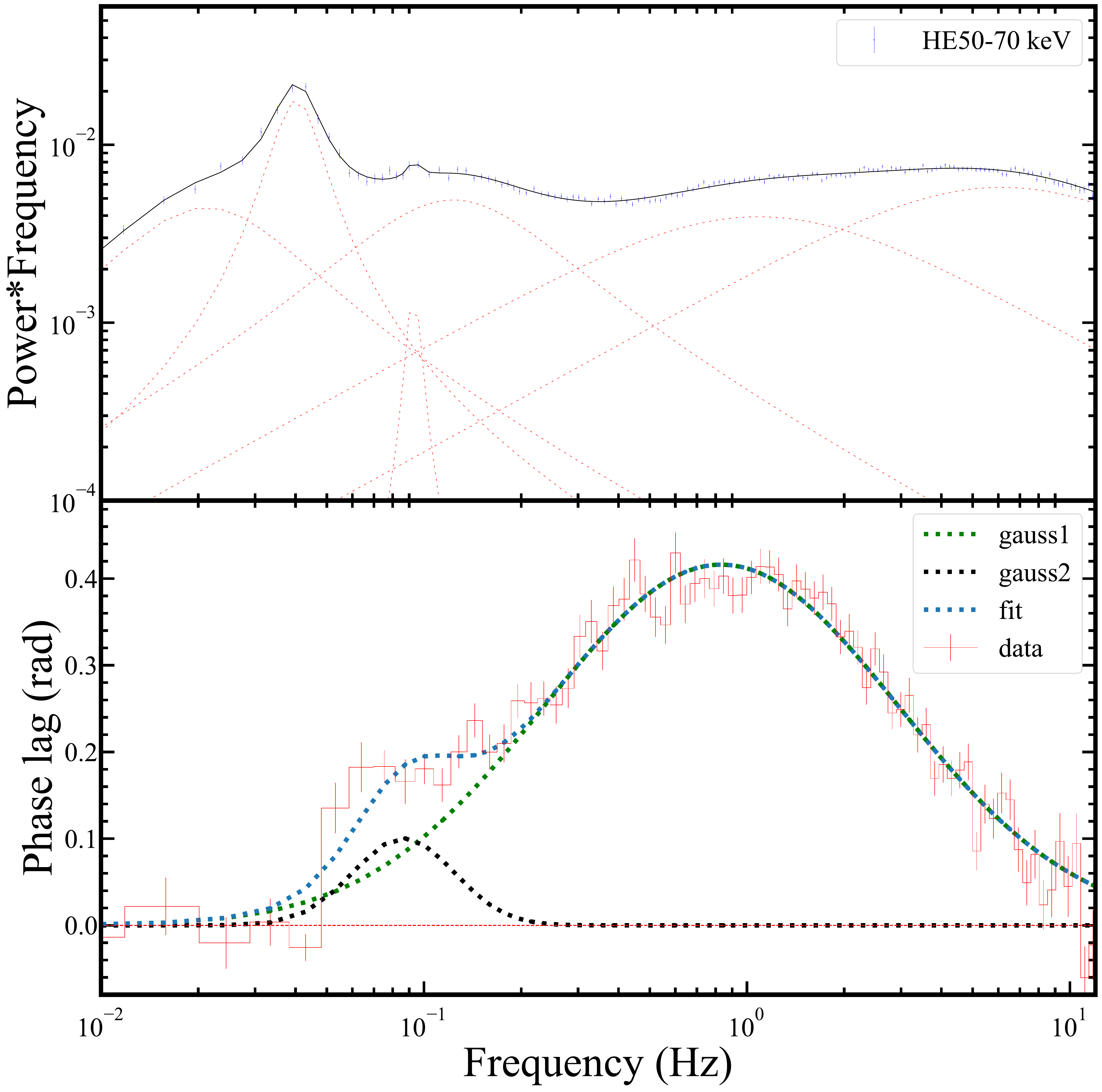}
    \caption{PDS for HE 50--70 keV and phase lag spectrum. The phase lag spectra was calculated for the LE 1--10 keV relative to the HE 50--70 keV. \emph{Insight}-HXMT ObsIDs P0114661003 and P0114661004 are used.}
    \label{pdslag}
\end{figure}

As shown in Figure~\ref{pdslag}, at frequencies above QPO, the value of phase lag is positive and frequency-dependent. The positive lag means that hard emission lags the soft one. According to fluctuation propagating model, hard photons coming from the inner region will lag behind soft photons coming from the outer region \citep{2001MNRAS.327..799K}. More interestingly, there are two humps in Figure~\ref{pdslag}. From Figure~\ref{pdslag}, we can approximately see that $\nu_{\rm g1} \simeq \nu_{\rm L3}$ and $\nu_{\rm g2} \simeq \nu_{\rm L2}$. This correspondence is very similar to the simulation results \citet{2017MNRAS.472.3821R} gave. According to the PROPFLUC model developed by \citet{2012MNRAS.419.2369I}, if we consider a disk+corona geometry, then we will also have two humps in phase lag spectra as two visible humps in PDS \citep{2014MNRAS.440.2882R,2016AN....337..524R,2016MNRAS.462.4078R,2017MNRAS.472.3821R,2017MNRAS.469.2011R}.

\begin{figure}
\epsscale{1.0}
\centering
	\includegraphics[width=0.48\textwidth]{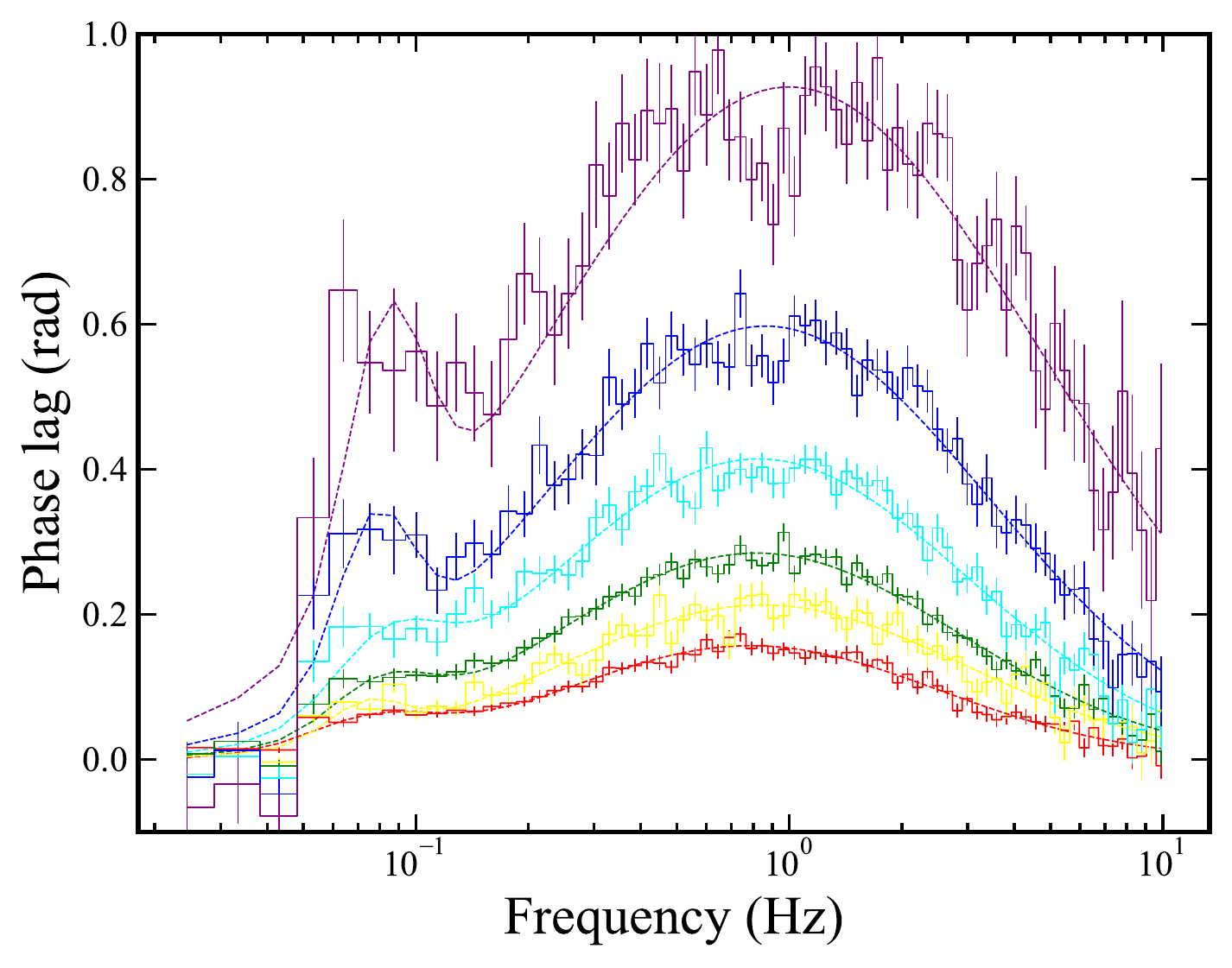}
    \caption{Phase lag spectrum for \emph{Insight}-HXMT ObsIDs P0114661003 and P0114661004. Reference energy band is 1--10 keV. From bottom to top is 10--20 keV (red), 20--30 keV (yellow), 30--50 keV (green), 50--70 keV (cyan), 70--100 keV (blue), 100--150 keV (purple) separately.}
    \label{lagm}
\end{figure}

In addition, as Figure~\ref{lagm} shows, the energy dependence of phase lag also reflects that harder photons come from the inner region and cause a greater delay \citep{2001MNRAS.327..799K}. 
 
In summary, by combining power spectra and phase lag spectra, we speculate that $L_{2}$ originates from the outer region whereas $L_{3}$ and $L_{4}$ originates from the inner region. Meanwhile, we confirm the applicability of fluctuation propagation model.

\begin{figure}
\epsscale{1.0}
\centering
	\includegraphics[width=0.48\textwidth]{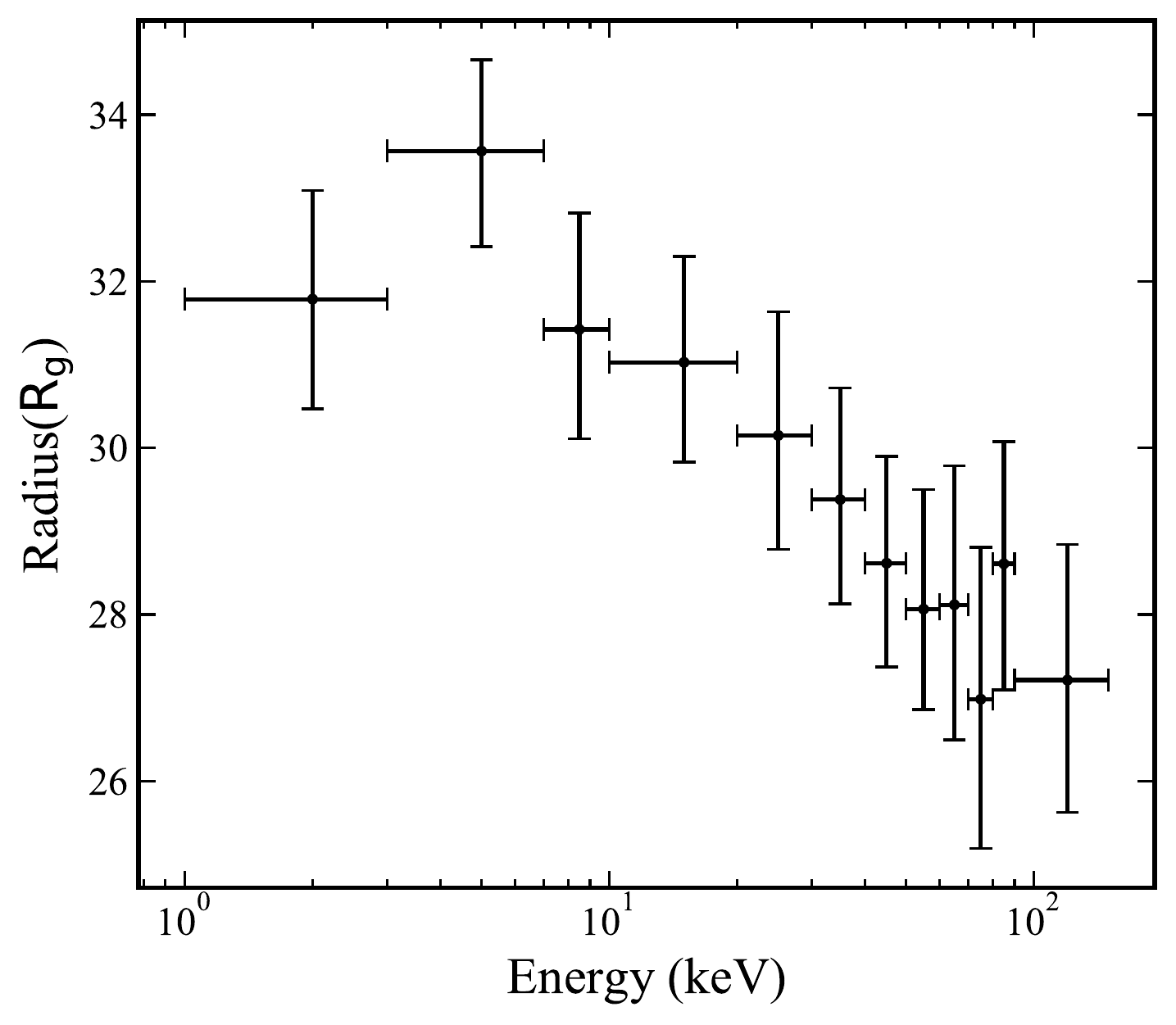}
    \caption{Truncation radius of accretion disk from the black hole as a function of photon energy for $L_2$ component. The standard $\alpha-$disk model was applied. A parameter set was used: $\alpha$=0.1, $\rm M=10\rm M_{\sun}$, scale height $H/r = 0.1$.}
    \label{R-E}
\end{figure}

\subsection{Implication for accretion structure}
In this paper, we investigate the accretion flow qualitatively at LE, ME, HE band and the quantitative fitting will be done in next paper preparing.
\citet{2021MNRAS.506.2020D} found significant spectral differences among Lorentzians for \target\ using the {\it NICER} 0.3--10 keV data. The model they presented comprises of an outer Comptonization region fueled by thermal photons from the cool disc, and an inner Comptonization region fueled by a fractional of the upscattered photons from the outer Comptonization region. Similarly, \citet{2022MNRAS.511..536K} also model the spectra and timing variability of MAXI J1820+070 to assume geometry consisting of variable disc and two hot flow regions. As for this paper, we also have found the two humps structure in PDS as papers above mentioned but one more QPO component in LE, ME and HE band. As \citet{2022MNRAS.511..536K} said, the dip in PDS is assumed to be caused by drop of viscous timescale between accretion disk and hot flow \citep{2017MNRAS.469.2011R,2022MNRAS.511..536K}. The different viscous timescale between the disc and
hot flow are physically natural because the scale height $H/R$ of the accretion
flow is expected to be different between these regions \citep{1997ApJ...476...49N}. Thus, we see two evident humps in PDS (separately $L_{2}$ and $L_3$, $L_{4}$).

In addition to the common evolutionary trend for $L_{2}$, $L_{3}$ and $L_{4}$, $L_{3}$ shows more sophisticated evolution trend. $L_{3}$ first appears in all three energy band in ObsID P0114661002, but as hardness decreases, $L_{3}$ disappears in LE band first, then disappears in ME band after nine ObsIDs ($\sim$ 18 days) too. This may implicate the change of emitting spectrum where $L_{3}$ originates: when hardness ratio is less than 0.61, the region represented by $L_3$ emits a relatively hard spectrum so that $L_3$ only appears in HE PDS. The distinct evolution trend of $L_{3}$ highlights the strong contrast with $L_{2}$ and $L_{4}$.

Then we investigate the hot accretion flow to explain other results especially the energy-dependence of $L_3$ and $L_4$. Similar to the argument \citet{2022MNRAS.511..536K} presented, we attribute $L_3$ to the outer corona while $L_4$ comes from the inner corona. The fluctuation propagation model predicts that the characteristic frequency of broadband noise is inversely proportional to radius. In view of the toy model we discussed, combining with the relation between characteristic frequency and photon energy, we speculate that the region farther from black hole in the outer corona has a relatively uniform distribution in parameters such as density and temperature to cause the constant frequency below 20--30 keV for $L_3$. Then when photon energy is greater than 20--30 keV, with increasing optical depth, the emitted spectrum becomes harder at a smaller radius to cause increasing characteristic frequency. This explanation also applies to $L_4$ only with more seed photon from the outer corona. Then we use the energy-dependence of characteristic frequency to make some simple quantitative estimation. We consider that the outer radius of hot flow $r_{\rm out}$ equals to the inner radius of accretion disk $r_{\rm disk} \sim 27R_g$ (see Figure~\ref{R-E}). It is difficult to connect the characteristic frequency with regions of the hot flow because of the poor understanding of hot flow now. Nevertheless, because the break frequency of broadband noise component is proportional to $r^{-3/2}$ in hot flow, then the inner radius of the outer corona is $r_{\rm in1} \sim 20R_g$, the inner radius of the inner corona is $r_{\rm in2} \sim 7R_g$. In calculation, we assume that the 90--150 keV radiation comes from the innermost region of each corona and 1--3 keV radiation comes from the outer region. This result is basically consistent with that in \citet{2022MNRAS.511..536K} and $R_{\rm in} \gtrsim 10 R_g$ in \citet{2021A&A...654A..14D}. Meanwhile, as Figure~\ref{lagm} shows, we found that $\nu_{\rm g1} \simeq \nu_{L3}$ and $\nu_{\rm g2} \simeq \nu_{L2}$. This phenomenon can be interpreted under fluctuation propagation model. This shows that the region where $L_{2}$ and $L_{3}$ originate really locates at outer region in accretion flow to cause two humps in phase lag spectra. As for the $L_{4}$ component, we should similarly see a hump in phase lag spectra whereas that was not the case. We consider that it is because the inner region emits mostly in the hard band and locates close to ISCO \citep{2017MNRAS.469.2011R}. 

All discussions above are based on the two coronae model. In fact, we note that there are still some results that are not easy to interpret.  For instance, in Figure~\ref{h-r},~\ref{h-f}, we find that the disappearance of $L_3$ in LE, ME energy band as spectra evolves. In the two coronae model, the emission region where $L_3$ represents locates between outer accretion disk and inner corona. It is relatively difficult to explain why $L_3$ emits so hard spectra  when hardness ratio is less than $\sim$ 0.61. Besides, from Figure~\ref{combine}, when photon energy is less than 20--30 keV, the characteristic frequency of $L_3$ and $L_3$ shows no evident change. It seems to be contrary to our usual understanding that harder photon comes from inner region. As a result, as \citet{2017MNRAS.469.2011R,2018MNRAS.474.2259M} and other paper discuss \citep{2021A&A...656A..63M}, we may consider more sophisticated accretion flow geometry (such as bending wave, viscous diffusion, outward fluctuation propagation effect, hot jet-emitting disk etc.) and multi-wavelength observations to explain all the results.
To sum up, MAXI J1820+070 is an ideal laboratory to study inhomogeneous stratified corona.

 Except the model we discuss in this paper, similar to the model that QPO comes from precession of jet \citep{2021NatAs...5...94M}, we can also attribute broadband noise to jet's contribution \citep{2005ApJ...635.1203M,2011ApJ...728...13N}. It should be note that the results in \citet{2021NatAs...5...94M} reveal the relationship between the jet precession and the LFQPO in the high energy band, jet precession model does not depend on whether the accretion disk is truncated. \citet{2013MNRAS.429L..20M} has proposed the internal shock model to consider the effect of fluctuation in accretion flow on jet ejecta. Especially for \citet{2020ApJ...896...33W}, by studying the relation between hard time lag in the high-frequency range, the high-frequency time lags are significantly correlated to the photon index derived from the fit to the quasi-simultaneous {\it NICER} spectrum. They suggested that this result is qualitatively consistent with a model in which the high-frequency time lags are produced by Comptonization in a jet. As \citet{2020ApJ...896...33W} shows, the evolution of the high-frequency lags is highly correlated to that of the photon index of hard spectral component by integrating the continuum broadband noise. Different from the method used in \citet{2020ApJ...896...33W}, we investigate the broadband noise components through Lorentzian function fitting method and mainly focus on the energy-dependence of each component in PDS.\  Overall, the results in \citet{2020ApJ...896...33W}, \citet{2021NatAs...5...94M} and our work
suggest that there are two hard emission regions during
the studied period of the outburst in MAXI J1820: one
is a hot flow liked corona and another one is a jet. As a result, this model needs further more investigation for connection between hot flow and jet base.

\section{CONCLUSION} \label{sec:highlight}





In summary, radial-stratified hot flow with truncated accretion disk is needed to explain our results based on fluctuation propagating model. We should combine timing analysis and spectral fitting especially in HE energy band to improve our understanding of the inhomogeneous corona in the future.  

\acknowledgments
We are grateful the anonymous referee's helpful comments and suggestions. This work has made use of the data from the \hxmt\ mission, a project funded by China National Space Administration (CNSA) and the Chinese
 Academy of Sciences (CAS), and data and/or software provided by the High Energy Astrophysics Science Archive Research Center (HEASARC), a service of
 the Astrophysics Science Division at NASA/GSFC. This
  work is supported by the National Key RD Program
 of China (2021YFA0718500) and the National Natural Science Foundation of China (NSFC) under grants
 U1838201, U1838202, 11733009, 11673023, U1938102, U2038104, U2031205, the CAS Pioneer Hundred Talent Program (grant No. Y8291130K2) and the Scientific and Technological innovation project of IHEP (grant No. Y7515570U1).


\textit{Softwares}: XSPEC \citep{1996ASPC..101...17A}, Astropy \citep{2013A&A...558A..33A}, Numpy \citep{2011CSE....13b..22V}, Matplotlib \citep{2007CSE.....9...90H}, Stingray \citep{2019ApJ...881...39H,2019JOSS....4.1393H}.
\section*{Data Availability}
The raw data underlying this article are available at http://hxmten.ihep.ac.cn/.

%

\vspace{5mm}

\bibliographystyle{aasjournal}
\bibliography{ms}{}



\end{document}